\documentclass[%
    reprint,superscriptaddress,
    nofootinbib,
    amsmath,amssymb,
    aps,pra,
    floatfix
]{revtex4-2}
\pdfoutput=1 

\usepackage{graphicx} 
\usepackage{dcolumn} 
\usepackage{xcolor}
\usepackage{bm} 
\usepackage{physics}
\usepackage{amssymb,amsmath,amsfonts,mathtools}
\usepackage{amsthm}
\usepackage{bbm}
\usepackage{algorithm}
\usepackage[rightComments=false]{algpseudocodex}
\usepackage{hyperref}
\hypersetup{
    colorlinks,
    citecolor=blue,
    filecolor=blue,
    linkcolor=blue,
    urlcolor=blue
}
\usepackage[]{mdframed}
\mdfsetup{innerbottommargin=\topskip}

\newcommand{\sgn}{\operatorname{sgn}}
\newcommand{\pos}{\operatorname{pos}}

\newtheorem{result}{Result}
\newtheorem{conjecture}{Conjecture}

\begin{document}

\title{Certification of genuine multipartite entanglement in spin ensembles with measurements of total angular momentum}

\author{Khoi-Nguyen Huynh-Vu}
\affiliation{College of Engineering and Computer Science, VinUniversity, Gia Lam district, Hanoi 14000, Vietnam}

\author{Lin Htoo Zaw}
\affiliation{Centre for Quantum Technologies, National University of Singapore, 3 Science Drive 2, Singapore 117543}

\author{Valerio Scarani}
\affiliation{Centre for Quantum Technologies, National University of Singapore, 3 Science Drive 2, Singapore 117543}
\affiliation{Department of Physics, National University of Singapore, 2 Science Drive 3, Singapore 117542} 

\begin{abstract}
We introduce entanglement witnesses for spin ensembles which detect genuine multipartite entanglement using only measurements of the total angular momentum. States that are missed by most other angular-momentum-based witnesses for spin ensembles, which include Greenberger-Horne-Zeilinger states and certain superpositions of Dicke states, can be effectively detected by our witness. The protocol involves estimating the probability that the total angular momentum is positive along equally-spaced directions on a plane. Alternatively, one could measure along a single direction at different times, under the assumption that the total spins undergo a uniform precession. Genuine multipartite entanglement is detected when the observed score exceeds a separable bound. Exact analytical expressions for the separable bound are obtained for spin ensembles $j_1\otimes j_2\otimes\dots \otimes j_N$ such that the total spin is a half-integer, and numerical results are reported for the other cases. Finally, we conjecture an expression for the separable bound when the total spin is not known, which is well supported by the numerical results.
\end{abstract}

\maketitle

\section{Introduction}
Entanglement is an important resource that enables many quantum protocols \cite{guhne_toth_2009,metrology-review,review-nature}. For this reason, the experimental certification of entanglement in real quantum devices is extremely important for both practical and foundational purposes. Consequently, many methods for detecting entanglement have been proposed and studied since the early days of quantum information \cite{horodecki_x4}. Entanglement witnesses can be constructed analytically or numerically \cite{sperling_ew_2013,spi}, and it is even possible to certify entanglement without full characterization of the measurement device \cite{witness-MDI,witness-random}.

However, there are some challenges when implementing entanglement witnesses in practice. Each experimental setup has measurements and operations that are native to that system. In general, arbitrarily constructed witnesses would require measurements in bases different from the native ones. So, some operations on the state need to be performed prior to measurement, which impacts the fidelity of the measurement. Alternatively, full or partial tomography must be done to extract the expectation value of the witness from the tomographic data, but tomography can be intractable for large systems. As such, when taking pragmatic reasons into account, witnesses that depend only on measurements native to the experimental setup might be more suitable in some cases.

In this paper, we focus on the \emph{spin ensemble}, a collection of particles labeled by $n$, each with a fixed spin $j_n$. Experimental settings in which spin ensembles appear include ultracold atoms in optical lattices \cite{optical-lattice-review}, and spin defects \cite{spin-defects-review} and donors \cite{spin-donors-review} in solid state materials. Measurements of angular momentum $\vec{J}^{(j_n)}$ are natural in such systems. We take that any component (e.g. $J_x$) or function of $\vec{J}$ (e.g. $\lvert\vec{J}\rvert^2$) can be easily measured.

Several novel witnesses that utilize only angular momentum measurements have been introduced for spin ensembles. A mainstay are the spin-squeezing inequalities \cite{spin_squeeze_extreme_2001,guhne_toth_2009, pezze_review_2018}, which are based on variances of different angular momentum components, and are built upon the uncertainty relations of spin observables. Following the seminal paper by \citet{sorensen_2001}, generalized spin-squeezing criteria have also been proposed to detect the entanglement of two to three spin-half particles \cite{korbicz_2005, korbicz_2006}, approximate many-body singlet states \cite{toth_singlets_2004}, and symmetric Dicke states \cite{toth_dicke_2007}. Families of such multipartite witnesses have also been characterized for spin-half particles \cite{toth_ssi_2007, toth_ssi_2009}, and later, for more general spin ensembles \cite{vitagliano_sse_2011, vitagliano_sse_2014}.

Another notable approach involves the energy observable, which, for spin ensembles, are some function of the angular momentum operators \cite{dowling_energy_2004}. If the ground state of the Hamiltonian is known to be entangled, and the measured energy is below a certain threshold, then the system itself must be entangled. Energy-based witnesses have been applied to various spin models, including $XY$ \cite{toth_spin_2005, giampaolo_gme_2013, igloi_toth_2023}, $XYZ$ \cite{homayoun_2019}, Heisenberg \cite{toth_spin_2005}, and more general spin chains \cite{wu_2005,troiani_2012,gabriel_2013}.

Since these witnesses act upon spin ensembles, the witnessed correlations are that of entanglement between many particles. Such multipartite entanglement takes on a more complex character in comparison to bipartite entanglement. First, the entanglement might be dependent on the partition chosen: For example, $\propto(\ket{\uparrow\downarrow}_{1,2}-\ket{\downarrow\uparrow}_{1,2})\otimes\ket{\downarrow}_3$ is entangled over the $\{1\}{\text{--}}\{2,3\}$ partition, but separable over the $\{1,2\}{\text{--}}\{3\}$ partition. As such, the stronger notion of \emph{genuine multipartite entanglement} (GME) has to be introduced: The entanglement present in a state is defined to be GME if it \emph{cannot} be written as a convex combination of separable states, where the separability of each state can be over \emph{any} bipartition \cite{GME-coined}. Second, there are inequivalent types of maximally entangled states that cannot be interconverted using only local operations and classical communication: For three spin-half particles, these are the $W$ state $\ket{W_3} \propto \ket{\uparrow\downarrow\downarrow} + \ket{\downarrow\uparrow\downarrow} + \ket{\downarrow\downarrow\uparrow}$ and the Greenberger-Horne-Zeilinger (GHZ) state $\ket{\text{GHZ}_3} \propto \ket{\uparrow}^{\otimes 3} + \ket{\downarrow}^{\otimes 3}$ \cite{inequivalent-tripartite-ent}.

In terms of the different types of multipartite entanglement, only some of the aforementioned angular-momentum-based witnesses can detect GME \cite{guhne_me_2005, guhne_toth_2006, giampaolo_gme_2013, teh_reid_2019, li_gme_2021}, and those that do are mostly effective at detecting Dicke states \cite{dicke_1954}, a generalization of $W$ states. Notably missing are angular-momentum-based witnesses that can detect $N$-partite GHZ states $\ket{\text{GHZ}_N} \propto \ket{\uparrow}^{\otimes N} + \ket{\downarrow}^{\otimes N}$ beyond the tripartite case. Since GHZ states are resources for distributed computing \cite{W-GHZ-resource}, and are also useful in quantum secret sharing protocols \cite{GHZ-secret}, angular-momentum-based witnesses that can detect GHZ states are desirable.

The primary contribution of this paper is to fill this gap. We introduce a witness of GME that requires only measurements of the total angular momentum of the spin ensemble, and can detect states that are missed by existing angular-momentum-based criteria, in particular, our witness can detect $N$-partite GHZ states.

\section{\label{sec:precession-protocol}The Precession Protocol}
\subsection{The Protocol as a nonclassicality test}
We first present the protocol as a nonclassicality test. It consists of many independent rounds. In each round, one system is prepared in some state, then its total angular momentum is measured along one of the directions
\begin{align}
    J_k
    &\coloneqq e^{-i (2\pi k/K) J_z/\hbar} J_x e^{i (2\pi k/K) J_z/\hbar} \nonumber \\
    &= \cos(2\pi k/K) J_x +
        \sin(2\pi k/K) J_y,
\end{align}
where $k\in\{0,1,\dots,K-1\}$, $K$ is a positive integer, and where the $x$--$y$ plane of the measurement directions can be chosen at one's convenience. After many such rounds, the average probability that $J_k$ was found to be positive is calculated as the score
\begin{equation}\label{eq:score}
    P_K \coloneqq \frac{1}{K} \sum_{k=0}^{K-1}\bqty{
        \Pr(J_k>0) +
        \frac{1}{2} \Pr(J_k = 0)
    }.
\end{equation}
The protocol does not specify, as conditions for its validity, that the state be the same in each round (of course, if the preparation is not under good control, the score will be low); nor does it matter that the measurement $k$ be chosen in any particular order or at random. However, the preparation of the state and the choice of direction $k$ for the measurement must be uncorrelated (we could say that ``the measurement should be random, i.e., unpredictable, from the point of view of the system''). 

This protocol is based on a test known as the ``precession protocol'' or ``Tsirelson protocol'' \cite{tsirelson_2006,Lin_2022,zaw2022dynamicsbased}, which is a witness of nonclassicality for single systems. The first expression comes from a dynamical assumption in those works that we do not use here (see Sec.~\ref{sss:dyn}); however, for the sake of a name, we shall continue to refer to this protocol as to \textit{the precession protocol}. 

The upper bound $P_K \leq \mathbf{P}_K^c$ predicted by classical theory is $\mathbf{P}_K^c = 1/2$ for $K$ even and $\mathbf{P}_K^c = (1+1/K)/2$ for $K$ odd \cite{tsirelson_2006,Lin_2022}. Meanwhile, the expected score for a quantum system in the state $\rho$ is given by $P_K = \tr(\rho Q_K)$, with
\begin{equation}\label{eq:QK-def}
    Q_K \coloneqq \frac{1}{K}\sum_{k=0}^{K-1}\pos(J_k).
\end{equation}
Here, $\pos(J_k)$ is defined on the eigenstates $\ket{j,m}_k$ of $J_k$, such that $J_k\ket{j,m}_k = \hbar m \ket{j,m}_k$ and $2\pos(J_k)\ket{j,m}_k = [1+\sgn(m)]\ket{j,m}_k$, with the usual convention $\sgn(0) = 0$.
There exist quantum states that violate the classical bound. In particular, the maximal eigenstates of $Q_K$ (the states that achieve the largest quantum score) are known in some cases \cite{Lin_2022}. Therefore, upon performing the protocol, the observation $P_K > \mathbf{P}_K^c$ detects the nonclassicality of the measured system.

\subsection{Protocol as an entanglement witness}
\begin{figure*}
    \centering
    \includegraphics{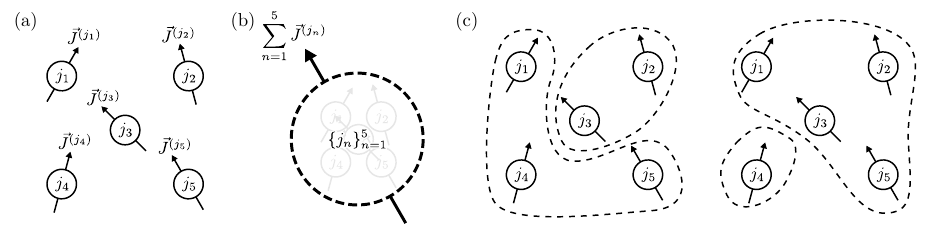}
    \caption{\label{fig:spin_ensemble}(a) We consider a spin ensemble of $N$ particles. Each particle has a fixed spin $j_n$ with angular momentum $\vec{J}^{(j_n)}$. (b) The protocol is performed on the total angular momentum $\vec{J} = \sum_{n=1}^N \vec{J}^{(j_n)}$ of the spin ensemble $\{j_n\}_{n=1}^N$, which is the sum of the individual angular momentum of each particle. (c) In general, the entanglement of a spin ensemble depends on how the ensemble is partitioned. The system might be entangled over one bipartition, but separable over another. As such, we are interested in the stronger notion of \emph{genuine multipartite entanglement} (GME): a state is GME if it cannot be written as a mixture of separable states, where the separability of the state can be over \emph{any} bipartition. Our entanglement witness detects GME if the score of the precession protocol is found to be larger than a separable bound.} 
\end{figure*}
We will now apply the precession protocol to the total angular momentum $\vec{J}$ of a composite system. Concretely, let us consider a spin ensemble consisting of $N$ particles with fixed spins $j_1,j_2,\dots,j_N$. With a slight abuse of notation, the angular momentum along the $k$ direction of the $n$th spin is denoted by
\begin{equation}
    J_k^{(j_n)} = \mathbbm{1}^{(j_1)} \otimes
        \dots \otimes
        \mathbbm{1}^{(j_{n-1})} \otimes
        J_{k}^{(j_n)} \otimes
        \mathbbm{1}^{(j_{n+1})} \otimes
        \dots \otimes
        \mathbbm{1}^{(j_N)},
\end{equation}
such that $J_k^{(j)}$ for a fixed spin is given by
\begin{equation}
    J_k^{(j)} = \sum_{m = -j}^{j} \hbar m  \ket{j,m}_k\hspace{-0.25em}\bra{j,m},
\end{equation}
where $\ket{j,m}_k$ is a simultaneous eigenstate of $\lvert{\vec{J}}\rvert^2$ and $J_k$ with eigenvalues $\hbar^2 j(j+1)$ and $\hbar m$ respectively. It is related to the usual eigenstate $\ket{j,m}$ of $\lvert{\vec{J}}\rvert^2$ and $J_z$ by $\ket{j,m}_k = e^{-i(2\pi k/K) J_z/\hbar}e^{-i(\pi/2) J_y/\hbar}\ket{j,m}$.  

With this, we can consider an implementation of the precession protocol with $J_k = \sum_{n=1}^N J_k^{(j_n)}$, and ask whether a composite system performs better in this protocol when the particles are entangled, in other words, whether we can define a \emph{precession-based entanglement witness (PEW)}. Previous works suggest it to be so: for two fixed spins, the state conjectured to be maximally violating is always entangled \cite{Lin_2022}; for two harmonic oscillators, violation by a linear combination of the original modes implies entanglement of the latter in a wide parameter range \cite{lin_pooja_2023}.

Indeed, we shall show that a sufficiently large violation of the classical bound by the spin ensemble implies GME. Our main results are as follows: an exact analytical bound when $\sum_{n=1}^N j_n = K/2$, for all odd $K\geq 3$ (Result \ref{thm:K-sep}); and a conjectured bound for all cases (Conjecture \ref{conj:K-sep}), with strong numerical evidence for $\sum_{n=1}^N j_n \leq 15$ and $K\leq 21$ (Result \ref{res:bi-K-sep}).

\subsection{Remarks on the implementation of the protocol}

\subsubsection{Collective measurements are possible but not necessary}

Our PEW belongs to the class of \textit{angular-momentum-based entanglement witnesses}. Indeed, it can be implemented by measuring only collective observables, namely, the suitable components of the total $\vec{J}$.

That being said, collective measurements are not necessary. Because $J_k = \sum_{n=1}^N J_k^{(j_n)}$, any measurement of $J_k$ can be done by first locally measuring $J_k^{(j_n)}$ for each spin, or $\sum_{n \in \{n_1,n_2,\dots\}} J_k^{(j_n)}$ for a subset of spins; then manually adding up the measurement outcomes afterwards. This feature is shared with some other witnesses \cite{toth_spin_2005,toth_dicke_2007}, but is contrasted with those that require measurements in a nonlocal basis \cite{toth_singlets_2004}.

\subsubsection{Measuring along only one direction under the assumption of dynamics}\label{sss:dyn}
In the earlier works about the precession protocol as a nonclassicality test \cite{tsirelson_2006,Lin_2022,zaw2022dynamicsbased}, the protocol was presented as a measurement of a single observable at different times, under the assumption that, during the certification process, that observable undergoes a uniform precession. Specifically, if $\vec{J} = (J_x,J_y,J_z)$ is known to precess uniformly about the $z$ axis with period $T\coloneqq 2\pi/\omega$ as
\begin{equation}
    \vec{J}(t) = \pmqty{
        \cos(\omega t)J_x(0) + \sin(\omega t)J_y(0)\\
        \cos(\omega t)J_y(0) - \sin(\omega t)J_x(0)\\
        J_z(0)
    },
\end{equation}
then indeed $J_k(0)=J_x(t_k)$ for $t_k=(k/K)T$. In this dynamical implementation, upon drawing $k$, one would wait until time $t_k$ and then measure $J_x$. In some systems, having to measure only in one direction could be an advantage, coming at the cost of the assumption on the dynamics. Hence, this offers another way of performing the precession protocol, rather than measuring at a fixed time along one among several directions as we present in this paper.

This alternate method can be useful in witnessing the entanglement of postquench states, which is a common problem in the study of spin ensembles \cite{igloi_toth_2023,quenchEW}. Here, ``\emph{quenching}'' refers to a sudden change of some parameters of the Hamiltonian. By starting with a highly entangling Hamiltonian, then quenching to the free Hamiltonian $H=-\omega J_z$, the dynamical assumption of uniform precession holds, and the precession protocol can be performed.

This approach fits naturally with recent efforts in optimal quantum control that seeks to engineer driven Hamiltonians that achieve target states with GME, like the GHZ state \cite{optimal-quantum-control-1,optimal-quantum-control-2}. Our protocol provides a way to certify the success of the quantum control procedure after it has been implemented.

\section{Main Results: Separable Bounds}\label{sec:results}
In this section, we detail the results with which we will eventually arrive at Results~\ref{thm:K-sep} and \ref{res:bi-K-sep} and Conjecture~\ref{conj:K-sep}.

Since we are interested in detecting GME states, we have to consider \emph{every possible bipartition} of the spin ensemble, as illustrated in Fig.~\ref{fig:spin_ensemble}. Let us therefore formalize the notion of a bipartition of a spin ensemble $\{j_n\}_{n=1}^N$. When partitioning the spin ensemble into two subsets, if the first subset is $\mathbf{J} = \{j_{n_1},j_{n_2},\dots,j_{n_L}\}$ containing $1 \leq L < N$ spins, then the second subset will contain the remaining spins $\mathbf{J}^\complement := \{j_n\}_{n=1}^N\setminus\mathbf{J}$. Meanwhile, for all possible bipartitions, the first subset can be any $\mathbf{J} \in 2^{\{j_n\}_{n=1}^N}\setminus\{ \varnothing, \{j_n\}_{n=1}^N\}$, where $2^{\{j_n\}_{n=1}^N}$ is the power set of $\{j_n\}_{n=1}^N$, and we exclude $\varnothing$ and its complement to ensure that at least one spin is present in each partition.

In addition, $\mathcal{H}^{(j_n)}$ denotes the Hilbert space spanned by $\{\ket{j_n,m}\}_{m=-j_n}^{j_n}$. We will also loosely call both $\sum_m \psi_m\ket{j_n,m}$ and $\sum_{m,m'} \rho_{m,m'} \ketbra{j_n,m}{j_n,m'}\succeq 0$, pure and mixed states, respectively, as ``states that belong in $\mathcal{H}^{(j_n)}$''.
 
With these notations, a state $\rho_{\mathbf{J},\mathbf{J}^\complement}$ of a spin ensemble is separable over the $\mathbf{J}$-$\mathbf{J}^\complement$ bipartition if $\rho_{\mathbf{J},\mathbf{J}^\complement} = \sum_k p_{k} \rho_{\mathbf{J},k} \otimes \rho_{\mathbf{J}^\complement,k}$, where $\rho_{\mathbf{J},k}$ (or $\rho_{\mathbf{J}^\complement,k}$) is a state within the subspace $\bigotimes_{j\in\mathbf{J}}\mathcal{H}^{(j)}$ (or $\bigotimes_{j'\in\mathbf{J}^\complement}\mathcal{H}^{(j')}$). Conversely, $\rho_{\text{GME}}$ is GME if it is not a convex combination of states separable over any bipartition $\mathbf{J}$: that is, $\rho_{\text{GME}} \neq \sum_{\mathbf{J}} p_{\mathbf{J}} \rho_{\mathbf{J},\mathbf{J}^\complement}$.

Finally, we define the separable bound $\mathbf{P}_K^{\text{sep}}(\{j_n\}_{n=1}^N)$ as the maximum score achieved by a non-GME state of the spin ensemble $\{j_n\}_{n=1}^N$, upon performing the precession protocol on the total angular momentum of the spin ensemble:
\begin{equation}\label{eq:GME-separable-bound}
\begin{aligned}
    \mathbf{P}_K^{\text{sep}}\pqty{\Bqty{j_n}_{n=1}^N}
    :\!\!&= \max_{\{p_{\mathbf{J}},\rho_{\mathbf{J},\mathbf{J}^\complement}\}} \tr[\pqty{\sum_{\mathbf{J}} p_\mathbf{J} \rho_{\mathbf{J},\mathbf{J}^\complement}}Q_K]\\
    &= \max_{\mathbf{J}} \max_{\rho_{\mathbf{J},\mathbf{J}^\complement}} \tr(\rho_{\mathbf{J},\mathbf{J}^\complement}Q_K).
\end{aligned}
\end{equation}
Here, convexity was used in the first line, $Q_K$ is as defined in Eq.~\eqref{eq:QK-def} with $\vec{J} = \sum_{n=1}^N\vec{J}^{(j_n)}$, while the maximizations are over all $\mathbf{J} \in 2^{\{j_n\}_{n=1}^N}\setminus\{ \varnothing, \{j_n\}_{n=1}^N\}$ and $\rho_{\mathbf{J},\mathbf{J}^\complement} = \sum_k p_k \rho_{\mathbf{J},k} \otimes \rho_{\mathbf{J}^\complement,k}$, where $\mathbf{J}^\complement = 2^{\{j_n\}_{n=1}^N}\setminus\mathbf{J}$.

By definition, every non-GME state must achieve the score $P_K \leq \mathbf{P}_K^{\text{sep}}(\{j_n\}_{n=1}^N)$. The negation of this statement implies that if the separable bound is violated, the system must be GME. Therefore, our PEW detects GME states if the score $P_K > \mathbf{P}_K^{\text{sep}}(\{j_n\}_{n=1}^N)$ is achieved when performing the precession protocol on the total angular momentum of the spin ensemble $\{j_n\}_{n=1}^N$.

\subsection{\label{sec:multi-to-bi}Relation between the multipartite and bipartite separable bounds}
By taking advantage of the decomposition of $Q_K$ in terms of the irreducible blocks of the angular momentum observable, the separable bounds of large spin ensembles can be related to separable bounds of two-spin systems.

For a particular bipartition $\mathbf{J} := \{j_{n_l}\}_{l=1}^L$, the usual rules for the addition of angular momentum applied to the total angular momentum $\vec{J} = \sum_{n=1}^N \vec{J}^{(j_n)}$ results in the decomposition \cite{angular-momentum-in-qm}
\begin{equation}\label{eq:multispin-decompose-J}
\begin{aligned}
    \vec{J} &= \sum_{l=1}^L \vec{J}^{(j_{n_l})} + \sum_{l'=L+1}^N \vec{J}^{(j_{n_{l'}})} \\[2ex]
    &= \bigoplus_{\tilde{\jmath}_1 = \abs{j_{n_1}-j_{n_2}}}^{j_{n_1} + j_{n_2}} 
    \;
    \bigoplus_{\tilde{\jmath}_2 = \abs{\tilde{\jmath}_1-j_{n_3}}}^{\tilde{\jmath}_1 + j_{n_3}}
    \dots
    \bigoplus_{\tilde{\jmath} = \abs{\tilde{\jmath}_{L-2}-j_{n_L}}}^{\tilde{\jmath}_{L-2} + j_{n_L}}
    \vec{J}^{(\tilde{\jmath})} \\
    &\qquad{}+{} \bigoplus_{\tilde{\jmath}'_1 = \abs{j_{n_{L+1}}-j_{n_{L+2}}}}^{j_{n_{L+1}} + j_{n_{L+2}}}
    \dots
    \bigoplus_{\tilde{\jmath}' = \abs{\tilde{\jmath}'_{N-L-2}-j_{n_N}}}^{\tilde{\jmath}_{N-L-2} + j_{n_N}}
    \vec{J}^{(\tilde{\jmath}')},\\[2ex]
    &\eqqcolon 
    \bigoplus_{\tilde{\jmath} \in \mathcal{J}(\mathbf{J})}
    \bigoplus_{\tilde{\jmath}' \in \mathcal{J}(\mathbf{J}^\complement)}
    \bqty{
        \vec{J}^{(\tilde{\jmath})}\otimes\mathbbm{1}^{(\tilde{\jmath}')} +
        \mathbbm{1}^{(\tilde{\jmath})}\otimes\vec{J}^{(\tilde{\jmath}')}
    },
\end{aligned}
\end{equation}
where $\mathcal{J}(\{j_{n_1},j_{n_2},\dots,j_{n_L}\}) := \{\tilde{\jmath} : 
\abs{\tilde{\jmath}_{L-2}-j_{n_L}} \leq \tilde{\jmath} \leq \tilde{\jmath}_{L-2}+j_{n_L}, 
\abs{\tilde{\jmath}_{L-3}-j_{n_{L-1}}} \leq \tilde{\jmath}_{L-2} \leq \tilde{\jmath}_{L-3}+j_{n_{L-1}},\dots,
\abs{j_{n_1}-j_{n_2}} \leq \tilde{\jmath}_1 \leq j_{n_1}+j_{n_2}
\}$ is the set of irreducible spins of the block decomposition of $\sum_{l=1}^L \vec{J}^{(j_{n_l})}$, which does not depend on the order of the elements of $\{j_{n_l}\}_{l=1}^L$ in its definition. The implicit tensor product with the identity has been made visible in Eq.~\eqref{eq:multispin-decompose-J}, which makes explicit that each term in the square brackets has nonzero support only in the $\mathcal{H}^{(\tilde{\jmath})}\otimes\mathcal{H}^{(\tilde{\jmath}')}$ subspace. 

It follows that, for an observable $f(\vec{J})$ that is a function of the total angular momentum,
\begin{equation}
\begin{aligned}
&\tr[ \rho_{\mathbf{J},\mathbf{J}^\complement} f(\vec{J}) ] \\
&\quad{}={}
    \!\!\!\!\!
    \sum_{
        \substack{
            \tilde{\jmath} \in \mathcal{J}(\mathbf{J}),\\
            \tilde{\jmath}' \in \mathcal{J}(\mathbf{J}^\complement)
        }
    }
    \!\!\!\!\!
    \tr[\rho_{\mathbf{J},\mathbf{J}^\complement} f\pqty{
        \mathbbm{1}^{(\tilde{\jmath})}\otimes\vec{J}^{(\tilde{\jmath}')} +
        \vec{J}^{(\tilde{\jmath})}\otimes\mathbbm{1}^{(\tilde{\jmath}')}
    }].
\end{aligned}
\end{equation}
In addition, using the convexity of the set $\{\rho_{\mathbf{J},\mathbf{J}^\complement}\}$,
\begin{equation}\label{eq:max-func-J}
\begin{aligned}
\max_{\rho_{\mathbf{J},\mathbf{J}^\complement}}\tr[\rho_{\mathbf{J},\mathbf{J}^\complement} f(\vec{J})] =
\max_{
    \lvert{\Psi_\mathbf{J},\Psi_{\mathbf{J}^\complement}}\rangle
} \langle{\Psi_\mathbf{J},\Psi_{\mathbf{J}^\complement}}\rvert f(\vec{J}) \lvert{\Psi_\mathbf{J},\Psi_{\mathbf{J}^\complement}}\rangle,
\end{aligned}
\end{equation}
where the maximization is over $\lvert{\Psi_\mathbf{J},\Psi_{\mathbf{J}^\complement}}\rangle = \lvert{\Psi_\mathbf{J}}\rangle \otimes \lvert{\Psi_{\mathbf{J}^\complement}}\rangle$, with $\lvert{\Psi_\mathbf{J}}\rangle \in \bigotimes_{j \in \mathbf{J}} \mathcal{H}^{(j)}$ and $\lvert{\Psi_{\mathbf{J}^\complement}}\rangle \in \bigotimes_{j' \in \mathbf{J}^\complement} \mathcal{H}^{(j')}$, which are pure states separable over the $\mathbf{J}$-$\mathbf{J}^\complement$ bipartition.

In relation to $\bigotimes_{j \in \mathbf{J}} \mathcal{H}^{(j)} = \bigoplus_{\tilde{\jmath} \in \mathcal{J}(\mathbf{J})} \mathcal{H}^{(\tilde{\jmath})}$, $\lvert{\Psi_\mathbf{J}}\rangle$ can be rewritten as $\lvert{\Psi_\mathbf{J}}\rangle = \bigoplus_{\tilde{\jmath} \in \mathcal{J}(\mathbf{J})} \sqrt{p_{\tilde{\jmath}}} \ket{\psi_{\tilde{\jmath}}}$ with independent parameters $\vec{p} = (p_{\tilde{\jmath}})_{\tilde{\jmath} \in \mathcal{J}(\mathbf{J})}$ and $\{\ket{\psi_{\tilde{\jmath}}}\}_{\tilde{\jmath} \in \mathcal{J}(\mathbf{J})}$, where $0 \leq p_{\tilde{\jmath}} \leq 1$, $\sum_{\tilde{\jmath} \in \mathcal{J}(\mathbf{J})} p_{\tilde{\jmath}} = 1$, and $\ket{\psi_{\tilde{\jmath}}} \in \mathcal{H}^{(\tilde{\jmath})}$.

Similarly with $\lvert{\Psi_{\mathbf{J}^\complement}}\rangle = \bigoplus_{\tilde{\jmath}' \in \mathcal{J}(\mathbf{J}^\complement)} \sqrt{p_{\tilde{\jmath}'}'} \ket{\psi_{\tilde{\jmath}'}}$, we have
\begin{equation}
\lvert{\Psi_\mathbf{J},\Psi_{\mathbf{J}^\complement}}\rangle = 
    \!\!\!\!\!
    \bigoplus_{
        \tilde{\jmath} \in \mathcal{J}(\mathbf{J}),
        \tilde{\jmath}' \in \mathcal{J}(\mathbf{J}^\complement)
    }
    \!\!\!\!\!
    \sqrt{p_{\tilde{\jmath}}p_{\tilde{\jmath}'}'} \ket{\psi_{\tilde{\jmath}},\psi_{\tilde{\jmath}'}},
\end{equation}
where $\ket{\psi_{\tilde{\jmath}},\psi_{\tilde{\jmath}'}} = \ket{\psi_{\tilde{\jmath}}}\otimes\ket{\psi_{\tilde{\jmath}'}}$. Therefore,
\begin{equation}
\begin{aligned}
    &\langle{\Psi_{\mathbf{J}},\Psi_{\mathbf{J}^\complement}}\rvert
    f\pqty{
        \mathbbm{1}^{(\tilde{\jmath})}\otimes\vec{J}^{(\tilde{\jmath}')} +
        \vec{J}^{(\tilde{\jmath})}\otimes\mathbbm{1}^{(\tilde{\jmath}')}
    }
    \lvert{\Psi_{\mathbf{J}},\Psi_{\mathbf{J}^\complement}}\rangle \\
    &\;{}={}
    p_{\tilde{\jmath}}p_{\tilde{\jmath}'}'
    \bra{\psi_{\tilde{\jmath}},\psi_{\tilde{\jmath}'}}
    f\pqty{
        \mathbbm{1}^{(\tilde{\jmath})}\otimes\vec{J}^{(\tilde{\jmath}')} +
        \vec{J}^{(\tilde{\jmath})}\otimes\mathbbm{1}^{(\tilde{\jmath}')}
    }
    \ket{\psi_{\tilde{\jmath}},\psi_{\tilde{\jmath}'}}.
\end{aligned}
\end{equation}
This parametrization of $\lvert{\Psi_{\mathbf{J}},\Psi_{\mathbf{J}^\complement}}\rangle$ in terms of $\vec{p}$, $\vec{p}'$, $\{\ket{\psi_{\tilde{\jmath}}}\}_{\tilde{\jmath} \in \mathcal{J}(\mathbf{J})}$, and $\{\ket{\psi_{\tilde{\jmath}'}}\}_{\tilde{\jmath}' \in \mathcal{J}(\mathbf{J}^\complement)}$ simplifies Eq.~\eqref{eq:max-func-J} to
\begin{widetext}
\begin{equation}\label{eq:max-func-J-multi-vs-bi}
\begin{aligned}
\max_{\rho_{\mathbf{J},\mathbf{J}^\complement}}\tr[\rho_{\mathbf{J},\mathbf{J}^\complement} f(\vec{J})]
&=
\max_{\vec{p},\vec{p}'} 
    \sum_{
        \tilde{\jmath} \in \mathcal{J}(\mathbf{J}),
        \tilde{\jmath}' \in \mathcal{J}(\mathbf{J}^\complement)
    } p_{\tilde{\jmath}}p_{\tilde{\jmath}'}' 
\bqty{
    \max_{\ket{\psi_{\tilde{\jmath}},\psi_{\tilde{\jmath}'}}}  \bra{\psi_{\tilde{\jmath}},\psi_{\tilde{\jmath}'}}
    f\pqty{
        \mathbbm{1}^{(\tilde{\jmath})}\otimes\vec{J}^{(\tilde{\jmath}')} +
        \vec{J}^{(\tilde{\jmath})}\otimes\mathbbm{1}^{(\tilde{\jmath}')}
    }
    \ket{\psi_{\tilde{\jmath}},\psi_{\tilde{\jmath}'}}
}\\
&= \max_{
    \tilde{\jmath} \in \mathcal{J}(\mathbf{J}),
    \tilde{\jmath}' \in \mathcal{J}(\mathbf{J}^\complement)
}
\bqty{
    \max_{\ket{\psi_{\tilde{\jmath}},\psi_{\tilde{\jmath}'}}}  \bra{\psi_{\tilde{\jmath}},\psi_{\tilde{\jmath}'}}
    f\pqty{
        \mathbbm{1}^{(\tilde{\jmath})}\otimes\vec{J}^{(\tilde{\jmath}')} +
        \vec{J}^{(\tilde{\jmath})}\otimes\mathbbm{1}^{(\tilde{\jmath}')}
    }
    \ket{\psi_{\tilde{\jmath}},\psi_{\tilde{\jmath}'}}
} \\
\therefore
\max_{\rho_{\mathbf{J},\mathbf{J}^\complement}}\tr[\rho_{\mathbf{J},\mathbf{J}^\complement}\; f\pqty{\sum_{j \in \mathbf{J}\cup\mathbf{J}^\complement}\vec{J}^{(j)}}]
&= \max_{
        \tilde{\jmath} \in \mathcal{J}(\mathbf{J}),
        \tilde{\jmath}' \in \mathcal{J}(\mathbf{J}^\complement)
}
\Bqty{
    \max_{\rho_{\{\tilde\jmath\},\{\tilde\jmath'\}}}\tr[\rho_{\{\tilde\jmath\},\{\tilde\jmath'\}}\; f\pqty{\sum_{j \in \{\tilde\jmath\}\cup\{\tilde\jmath'\}}\vec{J}^{(\tilde\jmath)}}]
},
\end{aligned}
\end{equation}
\end{widetext}
where the second line follows from the maximization of the convex sum in the first, and the last line similarly follows from the convexity of separable states $\rho_{\{\tilde\jmath\},\{\tilde\jmath'\}}$ in the subspace $\mathcal{H}^{(\tilde\jmath)}\otimes\mathcal{H}^{(\tilde\jmath')}$.

Notably, each term in the braces in Eq.~\eqref{eq:max-func-J-multi-vs-bi} is a maximization over separable states on the subspace of the tensor product of just the two fixed spins $\tilde{\jmath}$ and $\tilde{\jmath}'$, and the observable is itself defined within that subspace. With $f(\vec{J}) = Q_K$ in Eq.~\eqref{eq:max-func-J-multi-vs-bi} substituted into Eq.~\eqref{eq:GME-separable-bound}, the GME and bipartite separable bounds can be related as follows:
\begin{equation}\label{eq:multipartite-sep-bound}
    \mathbf{P}_K^{\text{sep}}\pqty{\Bqty{j_n}_{n=1}^N} = \max_{\mathbf{J}}
    \max_{
        \substack{
            \tilde{\jmath} \in \mathcal{J}(\mathbf{J}),\\
            \tilde{\jmath}' \in \mathcal{J}(\mathbf{J}^\complement)
        }
    }
    \mathbf{P}_K^{\text{sep}}\pqty{\{\tilde\jmath,\tilde\jmath'\}},
\end{equation}
with $\mathbf{J} \in 2^{\{j_n\}_{n=1}^N}\setminus\{ \varnothing, \{j_n\}_{n=1}^N\}$.

To evaluate $\mathbf{P}_K^{\text{sep}}({\{j_n\}_{n=1}^N})$, one first identifies the spins $\tilde\jmath$ and $\tilde\jmath'$ that appear in Eq.~\eqref{eq:multipartite-sep-bound}, evaluate $\mathbf{P}_k^{\text{sep}}\pqty{\{\tilde\jmath,\tilde\jmath'\}}$ for each pair, and pick out the maximum value. This simplifies the process of calculating the multipartite separable bound, as it reduces it to the calculation of several bipartite separable bounds.

A possible hassle might come from iterating over the possible subsets $\mathbf{J}$, or from the tedium of working out the set of irreducible spins $\mathcal{J}(\mathbf{J})$. Closed-form expressions of $\mathbf{J}$ and $\mathcal{J}(\mathbf{J})$ are known, and can be worked out algorithmically with a computer \cite{addition_same_spin}. Alternatively, from the necessity that $\tilde\jmath, \tilde\jmath'$ must be nonnegative, and that $\tilde\jmath + \tilde\jmath' \leq \sum_{n=1}^N j_n$ since they arise from the block decomposition of $\sum_{n=1}^N \vec{J}^{(j_n)}$, a convenient upper bound assuming that every bipartition has no spin-zero sectors in their respective partitions is
\begin{equation}\label{eq:multipartite-sep-upper-bound}
\begin{gathered}
    \text{given}\;\forall \mathbf{J} \in 2^{\{j_n\}_{n=1}^N} \setminus \{ \varnothing, \{j_n\}_{n=1}^N \} : 0 \notin \mathcal{J}(\mathbf{J})
    \cup
    \mathcal{J}(\mathbf{J}^{\complement}): \\
    \mathbf{P}_K^{\text{sep}}\pqty{\Bqty{j_n}_{n=1}^N} \leq
    \max_{\substack{
        0 \leq \tilde\jmath + \tilde\jmath' \leq \sum_{n=1}^N j_n,\\
        \tilde\jmath,\tilde\jmath' \in \frac{1}{2}\mathbb{Z}^+_0
    }}
    \mathbf{P}_K^{\text{sep}}\pqty{\{\tilde\jmath,\tilde\jmath'\}},
\end{gathered}
\end{equation}
where $\frac{1}{2}\mathbb{Z}^+_0$ is the set of positive integers and half-integers.

\subsection{Trivial separable bounds \texorpdfstring{for $\sum_{n=1}^N j_n <  K/2$}{when the total spin is less than half the number of measurements}}\label{sec:trivial-separable-bound}
For the special case $\sum_{n=1}^N j_n <  K/2$, the separable bound can be calculated directly. By performing a similar block decomposition to Eq.~\eqref{eq:multispin-decompose-J}, the total angular momentum of the spin ensemble is
\begin{equation}\label{eq:block-diagonal-J-multi}
    \vec{J} = \sum_{n=1}^N \vec{J}^{(j_n)} = \bigoplus_{j\in\mathcal{J}\pqty{\{j_n\}_{n=1}^N}}^{\sum_{n=1}^N j_n} \vec{J}^{(j)}.
\end{equation}
As before, the observable $Q_K$, which describes the precession protocol performed on the total angular momentum, will also inherit this block diagonal structure:
\begin{equation}\label{eq:QK-block-full}
    Q_K 
    = \bigoplus_{j\in\mathcal{J}\pqty{\{j_n\}_{n=1}^N}}^{\sum_{n=1}^N j_n}   \underbrace{\frac{1}{K}\sum_{k=0}^{K-1}\pos[J_k^{(j)}]}_{\eqqcolon Q_K^{(j)}}.
\end{equation}
The eigendecomposition of $Q_K^{(j)}$ is known for some values of $j$ \cite{Lin_2022}. In particular,
\begin{equation}\label{eq:QK, d < K+1}
    Q_K^{(j<K/2)} = \frac{1}{2}\mathbbm{1}^{(j)}.
\end{equation}
Since it must be that $j \leq \sum_{n=1}^N j_n$ for every $j\in\mathcal{J}({\{j_n\}_{n=1}^N})$,
\begin{equation}
    Q_K
    = \bigoplus_{j\in\mathcal{J}\pqty{\{j_n\}_{n=1}^N}}^{\sum_{n=1}^N j_n < K/2}
    \frac{1}{2}\mathbbm{1}^{(j)}
    = \frac{1}{2} \mathbbm{1}.
\end{equation}
Hence, for a spin ensemble such that $\sum_{n=1}^N j_n< K/2$, $P_K = 1/2$ for \emph{every} state of the system. This also trivially gives $\mathbf{P}_K^{\text{sep}}(\{j_n\}_{n=1}^N) = 1/2$, but since there are no states that can violate this separable bound, performing the precession protocol in this case does not reveal anything about the entanglement of the system.

\subsection{Separable bounds \texorpdfstring{for $\sum_{n=1}^N j_n =  K/2$}{when the total spin is equal to half the number of measurements}}\label{sec:sep-bound-K/2}
For $\sum_{n=1}^N j_n =  K/2$, the upper bound in Eq.~\eqref{eq:multipartite-sep-upper-bound} requires the values of $\mathbf{P}_K^{\text{sep}}(\{\tilde\jmath,\tilde\jmath'\})$ for every $\tilde\jmath$ and $\tilde\jmath'$ such that $\tilde\jmath+\tilde\jmath' \leq K/2$. Since the separable bounds for $\tilde\jmath+\tilde\jmath' < K/2$ have already been evaluated in the previous section, we would only need to additionally work out $\mathbf{P}_K^{\text{sep}}(\{\tilde\jmath,\tilde\jmath'\})$ for $\tilde\jmath+\tilde\jmath' = K/2$.

Hence, it is adequate to first restrict ourselves to the $\mathcal{H}^{(\tilde\jmath)}\otimes\mathcal{H}^{(\tilde\jmath')}$ subspace. A block decomposition similar to the one in Eq.~\eqref{eq:QK-block-full} gives
\begin{equation}\label{QK-pre-eigen-decomposition}
    Q_K
    = \!\!\!\!\!\bigoplus_{j=\abs{\tilde\jmath-\tilde\jmath'}}^{\tilde\jmath+\tilde\jmath'=K/2}\!\!\!\!\! Q_K^{(j)}
    = \frac{1}{2}\pqty{\mathbbm{1} - \mathbbm{1}^{(j=K/2)}} \oplus Q_K^{(j=K/2)},
\end{equation}
where we used $Q_K^{(j<K/2)} = \mathbbm{1}^{(j<K/2)}/2$. However, the eigendecomposition of $Q_K^{(j=K/2)}$ is \cite{Lin_2022}
\begin{equation}
\begin{aligned}
    Q_K^{(j=K/2)} &= \frac{1}{2}\mathbbm{1}^{(j=K/2)} + \frac{2^{-(K-1)}}{2}\pmqty{K-1\\\frac{K-1}{2}}\\
    &\qquad{}\times{}\pqty\Big{
        \ketbra{\mathbf{P}_{+K}} - 
        \ketbra{\mathbf{P}_{-K}}
    },
\end{aligned}
\end{equation}
where $\ket{\mathbf{P}_{\pm K}} = (\ket{\frac{K}{2},\frac{K}{2}} \pm (-1)^{(K-1)/2}\ket{\frac{K}{2},-\frac{K}{2}})/\sqrt{2}$. Placing this back into Eq.~\eqref{QK-pre-eigen-decomposition},
\begin{equation}\label{eq:QK-eigen-decomposition}
\begin{aligned}
    Q_K &= \frac{1}{2}\mathbbm{1} + \frac{2^{-(K-1)}}{2}\pmqty{K-1\\\frac{K-1}{2}}\\
    &\qquad{}\times{}\pqty\Big{
        \ketbra{\mathbf{P}_{+K}} - 
        \ketbra{\mathbf{P}_{-K}}
    }.
\end{aligned}
\end{equation}

Meanwhile, observing that the only way for $-\tilde\jmath \leq \tilde m \leq \tilde\jmath$ and $-\tilde\jmath' \leq \tilde m' \leq \tilde\jmath'$ to satisfy $\tilde m + \tilde m' = \pm K/2$ is to have $\tilde m = \pm \tilde\jmath$ and $\tilde m' = \pm \tilde\jmath'$, $\ket{\mathbf{P}_{\pm K}}$ is expressed in the $\{ \ket{\tilde\jmath,\tilde m} \otimes \ket{\tilde\jmath',\tilde m'} \}_{\tilde m, \tilde m' = -\tilde\jmath,-\tilde\jmath'}^{\tilde\jmath,\tilde\jmath'}$ basis as
\begin{equation}
\begin{aligned}\label{eq:states-pmK}
    \ket{\mathbf{P}_{\pm K}} = &\frac{1}{\sqrt{2}}\Big(
        \ket{\tilde\jmath,\tilde\jmath}\otimes\ket{\tilde\jmath',\tilde\jmath'}\\ &\quad\qquad{}\pm{}
        (-1)^{(K-1)/2}
        \ket{\tilde\jmath,-\tilde\jmath}\otimes\ket{\tilde\jmath',-\tilde\jmath'}
    \Big).
\end{aligned}
\end{equation}
With this, we are now able to evaluate
\begin{equation}\label{eq:Psep,j1+j2=K/2}
    \mathbf{P}_K^{\text{sep}}(\{\tilde\jmath,\tilde\jmath'\}) = \max_{\!\lvert{\psi_{\tilde\jmath},\psi_{\tilde\jmath'}}\rangle\!} \langle {\psi_{\tilde\jmath},\psi_{\tilde\jmath'}}\rvert Q_K \lvert{\psi_{\tilde\jmath},\psi_{\tilde\jmath'}}\rangle,
\end{equation}
where $\lvert{\psi_{\tilde\jmath},\psi_{\tilde\jmath'}}\rangle = \lvert{\psi_{\tilde\jmath}}\rangle\otimes\lvert{\psi_{\tilde\jmath'}}\rangle$ with $\lvert\psi_{\tilde\jmath}\rangle \in \mathcal{H}^{(\tilde\jmath)}$ and $\lvert\psi_{\tilde\jmath'}\rangle \in \mathcal{H}^{(\tilde\jmath')}$. Let us first parametrize the latter state as $\lvert{\psi_{\tilde\jmath'}}\rangle = \sum_{\tilde m'=-\tilde\jmath'}^{\tilde\jmath'} c_{\tilde m'} \ket{\tilde\jmath',\tilde m'}$. Then, 
\begin{equation}\label{eq:separable-operator}
\begin{aligned}
    &\pqty{\mathbbm{1}^{(\tilde\jmath)}\otimes\langle{\psi_{\tilde\jmath'}}\rvert}Q_K\pqty{\mathbbm{1}^{(\tilde\jmath)}\otimes\lvert{\psi_{\tilde\jmath'}}\rangle}\\
    &\quad\qquad{}\widehat{=}{} \frac{1}{2}\mathbbm{1}^{(\tilde\jmath)} + (-1)^{\frac{K-1}{2}}\frac{2^{-(K-1)}}{4}\pmqty{K-1\\\frac{K-1}{2}}\\
    &\quad\qquad\qquad{}\times{} \pmqty{
        \abs{c_{-j}}^2 & 0 & \hdots & 0 & c_{-j}^* c_{+j}\\
        0 & 0 & \hdots & 0 & 0 \\
        \vdots &  & \ddots & & \vdots \\
        0 & 0 & \hdots & 0 & 0 \\
        c_{-j} c_{+j}^* & 0 & \hdots & 0 & \abs{c_{+j}}^2
    },
\end{aligned}
\end{equation}
where the operator is represented as a matrix in the $\{\ket{\tilde\jmath,\tilde m}\}_{\tilde m=-\tilde\jmath}^{\tilde\jmath}$ basis. Since $\max_{\lvert{\psi_{\tilde\jmath}}\rangle} \langle {\psi_{\tilde\jmath},\psi_{\tilde\jmath'}}\rvert Q_K \lvert{\psi_{\tilde\jmath},\psi_{\tilde\jmath'}}\rangle$ \emph{for a fixed} $\lvert{\psi_{\tilde\jmath'}}\rangle$ is the largest eigenvalue of the operator in Eq.~\eqref{eq:separable-operator}, Eq.~\eqref{eq:Psep,j1+j2=K/2} can be evaluated by first diagonalizing Eq.~\eqref{eq:separable-operator} using standard analytical methods, then maximizing over $\lvert{\psi_{\tilde\jmath'}}\rangle$. Therefore,
\begin{equation}\label{eq:K-sep}
\begin{aligned}
    \mathbf{P}_K^{K\text{-sep}}
    :\!\!&= \mathbf{P}_K^{\text{sep}}(\{\tilde\jmath,\tilde\jmath'\})\;\,\text{for $\tilde\jmath+\tilde\jmath' = K/2$}\\
    &=
    \max_{\lvert{\psi_{\tilde\jmath'}}\rangle} \pqty{ \max_{\lvert{\psi_{\tilde\jmath}}\rangle} \langle {\psi_{\tilde\jmath},\psi_{\tilde\jmath'}}\rvert Q_K \lvert{\psi_{\tilde\jmath},\psi_{\tilde\jmath'}}\rangle } \\
    &= \!\!\max_{\{c_{\tilde m'}\}_{\tilde m'}}\!\! \bqty{\tfrac{1}{2} + \tfrac{2^{-(K-1)}}{4}\spmqty{K-1\\\frac{K-1}{2}}\pqty{\abs{c_{-\tilde\jmath'}}^2+\abs{c_{+\tilde\jmath'}}^2}} \\
    &= \frac{1}{2}\bqty{1 + 2^{-K}\pmqty{K-1\\\frac{K-1}{2}}}.
\end{aligned}
\end{equation}
Turning our attention back to the spin ensemble $\{j_n\}_{n=1}^N$ such that $\sum_{n=1}^N j_n =  K/2$, Eq.~\eqref{eq:multipartite-sep-upper-bound} gives
\begin{equation}\label{eq:multipartite-K-sep}
    \mathbf{P}_K^{\text{sep}}\pqty{\Bqty{j_n}_{n=1}^N} \leq \max\Bqty{1/2,\mathbf{P}_K^{K\text{-sep}}} = \mathbf{P}_K^{K\text{-sep}},
\end{equation}
where $1/2$ appears in the maximization from the contribution of $\mathbf{P}_K^{\text{sep}}(\tilde\jmath,\tilde\jmath')$ for the case $\tilde\jmath+\tilde\jmath' < K/2$, while $\mathbf{P}_K^{K\text{-sep}}$ covers the $\tilde\jmath+\tilde\jmath' = K/2$ case.

From Eq.~\eqref{eq:multipartite-K-sep}, Result~\ref{thm:K-sep} follows immediately.

\begin{mdframed}
\begin{result}\label{thm:K-sep}
Consider an $N$-partite spin ensemble $\{j_n\}_{n=1}^{N}$ such that $\sum_{n=1}^N j_n = K/2$ for odd $K \geq 3$. Perform the precession protocol with $K$ measurements on the total angular momentum of the system. If the score $P_K > \mathbf{P}_K^{K\text{\normalfont-sep}}$ is obtained, where
\begin{equation}
    \mathbf{P}_K^{K\text{\normalfont-sep}} = \frac{1}{2}\bqty{1 + 2^{-K}\spmqty{K-1\\\frac{K-1}{2}}},
\end{equation}
the spin ensemble is GME.
\end{result}
\end{mdframed}
Note that $\mathbf{P}_{K\leq 5}^{K\text{\normalfont-sep}} < \mathbf{P}_K^c$ and $\mathbf{P}_{K > 5}^{K\text{\normalfont-sep}} > \mathbf{P}_K^c$. That is, violating the classical bound is sufficient for detecting entanglement when $\sum_{n=1}^N j_n \leq 5/2$, while a larger violation is required when $\sum_{n=1}^N j_n > 5/2$. Regardless, the violation only needs to be, at most, $\sim 3.9\%$ larger than the classical bound in the latter case: a loose upper bound is $\mathbf{P}_{K}^{K\text{\normalfont-sep}} < 1.0391\mathbf{P}_K^c$ for all $K$.

\subsection{Separable bounds for general \texorpdfstring{$\sum_{n=1}^N j_n$}{total spin}, fixed \texorpdfstring{$K$}{number of measurements}}
For spin ensembles with general $\sum_{n=1}^N j_n$, the values of $\mathbf{P}_K^{\text{sep}}(\{\tilde\jmath,\tilde\jmath'\})$ for $\tilde\jmath+\tilde\jmath' > K/2$ also come into play in Eq.~\eqref{eq:multipartite-sep-bound}. However, these bipartite separable bounds cannot be solved analytically, and we have to instead rely on numerical methods.

Two numerical methods were used to evaluate the bipartite separable bounds: the first is a variant of separability power iteration, which provides lower bounds \cite{spi}; the second is semidefinite programming, whose global convergence is guaranteed, which provides conservative upper bounds \cite{boyd_1996}. Note that the latter is found by maximizing over a superset of the set of separable states that includes bound entangled states \cite{bound-entanglement,bound-entanglement-GME}, so the upper bounds might be loose in general.

Implementation of the numerical methods are detailed in Appendix~\ref{appendix: numerical methods}, while the scripts and generated data are available in Ref.~\cite{scripts}. By using both techniques, we are able to ascertain that the true value $\mathbf{P}_K^{\text{sep}*}$ falls within a range $\mathbf{P}_K^{\text{sep}} - \delta\mathbf{P}_K^{\text{sep}} \leq \mathbf{P}_K^{\text{sep}*} \leq \mathbf{P}_K^{\text{sep}} + \delta\mathbf{P}_K^{\text{sep}}$. The numerical errors $\delta\mathbf{P}_K^{\text{sep}}$ are plotted in Appendix~\ref{apd:extra-figs}.

\subsubsection{Numerical results for general \texorpdfstring{$\{\tilde\jmath,\tilde\jmath'\}$}{spin pair}, fixed \texorpdfstring{$K$}{number of measurements}}
\begin{figure}[ht]
    \centering
    \includegraphics{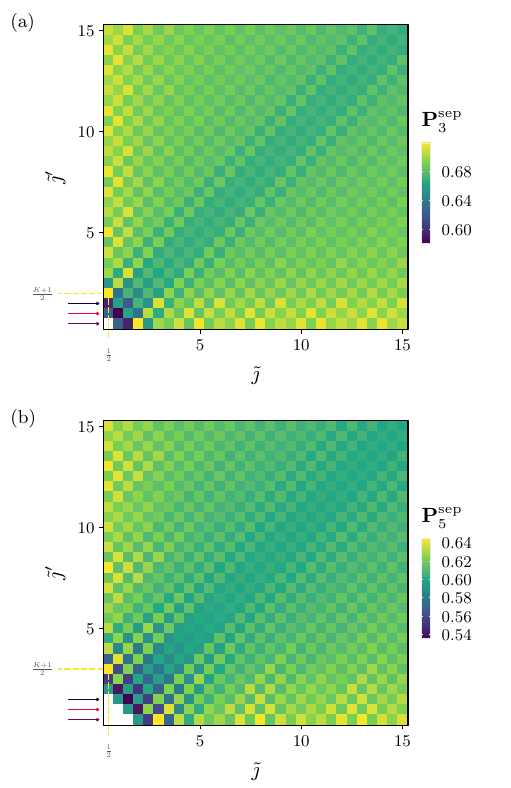}
    \caption{\label{fig:heatmap_K3_K5}Heatmap of $\mathbf{P}_{K}^{\text{sep}}(\{\tilde\jmath, \tilde\jmath' \})$ against $\tilde\jmath$ and $\tilde\jmath'$, for (a) $K=3$ and (b) $K=5$. Note that the values $\mathbf{P}_{K}^{\text{sep}}(\{\tilde\jmath, \tilde\jmath' \}) = 1/2$ for $\tilde\jmath + \tilde\jmath' < K/2$ are neither plotted here nor for the subsequent figures. The separable bounds are large when $\min(\tilde\jmath,\tilde\jmath')$ is small, and $\mathbf{P}_{K}^{\text{sep}}$ decreases as $\min(\tilde\jmath,\tilde\jmath')$ increases. The maximum value of $\mathbf{P}_{K}^{\text{sep}}$ in these cases occur at $\{\tilde\jmath,\tilde\jmath'\} = \{1/2,(K+1)/2\}$. The colored arrows mark out the direction along which the line cuts in Fig.~\ref{fig:linecut_K3_K5} are taken.}
\end{figure}

The plots of $\mathbf{P}_{K}^{\text{sep}}(\{\tilde\jmath,\tilde\jmath'\})$ against $\tilde\jmath$ and $\tilde\jmath'$ are shown in Fig.~\ref{fig:heatmap_K3_K5} for $K=3$ and $K=5$. Notice that the separable bound is large for $\min(\tilde\jmath,\tilde\jmath') < 2$, and takes on smaller values as $\min(\tilde\jmath,\tilde\jmath')$ increases.

\begin{figure}[ht]
    \centering
    \includegraphics{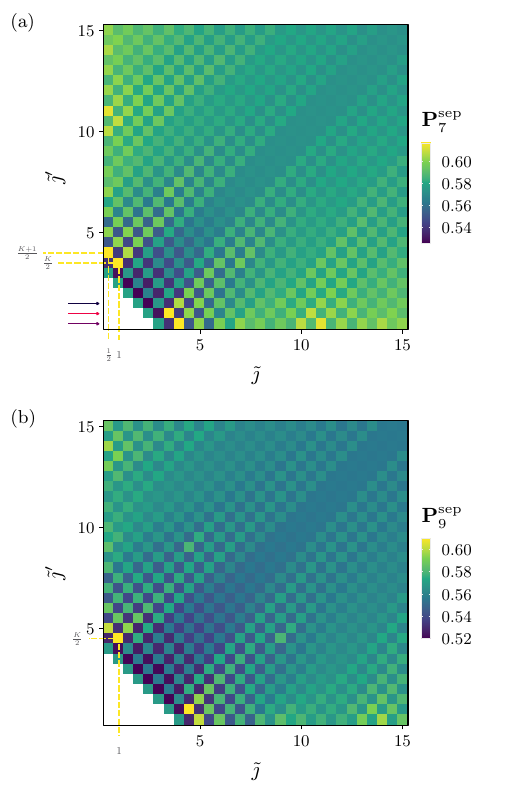}
    \caption{\label{fig:heatmap_K7_K9}Heatmap of $\mathbf{P}_{K}^{\text{sep}}(\{\tilde\jmath, \tilde\jmath' \})$ against $\tilde\jmath$ and $\tilde\jmath'$, for (a) $K=7$ and (b) $K=9$. Similarly to Fig.~\ref{fig:heatmap_K3_K5}, the separable bounds are large for small values of $\min(\tilde\jmath,\tilde\jmath')$. For $K=7$, $\mathbf{P}_{K}^{\text{sep}}$ is maximal at both $\{\tilde\jmath,\tilde\jmath'\} = \{1/2,(K+1)/2\}$ and $\{1,K/2\}$. For $K=9$, the maximum value occurs at $\{\tilde\jmath,\tilde\jmath'\} = \{1,K/2\}$. The colored arrows mark out the line cuts plotted in Fig.~\ref{fig:linecut_K7}.}
\end{figure}

This behavior [that the separable bound is large when $\min(\tilde\jmath$, $\tilde\jmath')$ is small] can be found not just for $K=7$ and $K=9$, as plotted in Fig.~\ref{fig:heatmap_K7_K9}, but up to $K=21$, as plotted in Fig.~\ref{fig:heatmap_big}. From the numerical results, we can find the values of $\{\tilde\jmath,\tilde\jmath'\}$ where the maximum occurs. As the gap between the largest and second largest value is larger than the numerical error, we can be certain that the largest computed value is indeed the maximum, leading us to the following result.
\begin{result}\label{res:sep-argmax}
The maximum separable bound over $0 \leq \tilde\jmath,\tilde\jmath' \leq 15$ for fixed $3 \leq K \leq 21$ occurs at
\begin{equation}\label{eq:res-argmax}
\underset{\{\tilde\jmath,\tilde\jmath'\}}{\operatorname{argmax}}\, \mathbf{P}_K^{\text{\normalfont sep}}(\{\tilde\jmath,\tilde\jmath'\}) =
\begin{cases}
    \Bqty{\frac{1}{2},\frac{K+1}{2}} & \text{for $K \leq 7$,} \\[1ex]
    \Bqty{1,\frac{K}{2}} & \text{for $K \geq 7$,}
\end{cases}
\end{equation}
where $\operatorname{argmax}_x f(x) = \mathbf{x}$ such that $f(\mathbf{x}) = \max_x f(x)$.
\end{result}
Equation~\eqref{eq:res-argmax} is multivalued for $K=7$ because, up to numerical precision, $\mathbf{P}_7^{\text{sep}}(\{1/2,4\}) = \mathbf{P}_7^{\text{sep}}(\{1,7/2\})$.

The numerical results for various values of $K$, especially when placed side-by-side as in Fig.~\ref{fig:heatmap_big}, strongly suggest that these behaviors should hold for any value of $K$. As such, for a given $K$, it is worth identifying some trends about the values of $\{\tilde\jmath,\tilde\jmath'\}$ at which the maximum separable bound occurs.

\begin{figure}[ht]
    \centering
    \includegraphics{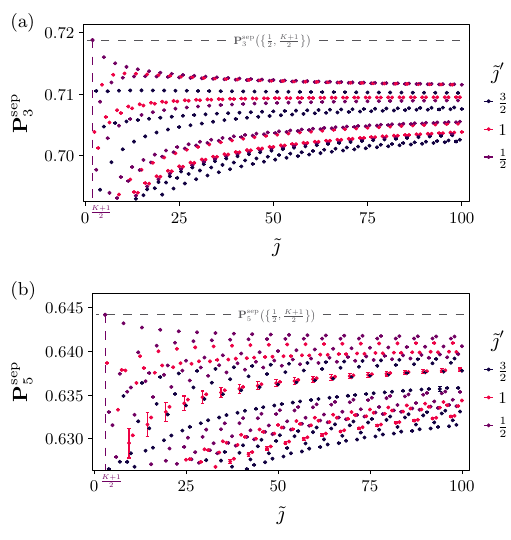}
    \caption{\label{fig:linecut_K3_K5}Line cuts from Fig.~\ref{fig:heatmap_K3_K5} of (a) $\mathbf{P}_3^{\text{sep}}(\{\tilde\jmath,\tilde\jmath'\})$ and (b) $\mathbf{P}_5^{\text{sep}}(\{\tilde\jmath,\tilde\jmath'\})$ along fixed values of $\tilde\jmath'$, plotted against a larger range of $\tilde\jmath$. For this and subsequent line cuts, separable bounds for $\tilde\jmath < \tilde\jmath'$ are not shown; they can be obtained by swapping $\tilde\jmath \leftrightarrow \tilde\jmath'$. Note also that error bars $\pm\delta \mathbf{P}_K^{\text{sep}}(\{\tilde\jmath,\tilde\jmath'\})$ are drawn for every point, but most are too small to be visible. As $\tilde\jmath$ increases, $\mathbf{P}_K^{\text{sep}}$ clusters around a range of values that are strictly smaller than $\mathbf{P}_K^{\text{sep}}(\{1/2,(K+1)/2\})$. Due to this converging behavior for large $\tilde\jmath$, we conjecture that the maximum value of $\mathbf{P}_K^{\text{sep}}(\{\tilde\jmath,\tilde\jmath'\})$ over $0 \leq \tilde\jmath,\tilde\jmath' \leq \infty$ occurs at $\{\tilde\jmath,\tilde\jmath'\} = \{1/2,(K+1)/2\}$ for $K=3$ and $K=5$.}
\end{figure}

\begin{figure}[ht]
    \centering
    \includegraphics{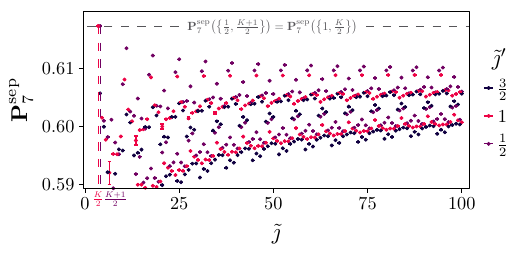}
    \caption{\label{fig:linecut_K7}Line cuts from Fig.~\ref{fig:heatmap_K7_K9}(a) of $\mathbf{P}_7^{\text{sep}}(\{\tilde\jmath,\tilde\jmath'\})$ along fixed values of $\tilde\jmath'$. Error bars $\pm\delta \mathbf{P}_7^{\text{sep}}(\{\tilde\jmath,\tilde\jmath'\})$ have been drawn for every point, but most are too small to be visible. Up to numerical precision, $\mathbf{P}_7^{\text{sep}}(\{1/2,4\}) = \mathbf{P}_7^{\text{sep}}(\{1,7/2\})$,
    and $\mathbf{P}_7^{\text{sep}}$ clusters around a range strictly smaller than $\mathbf{P}_7^{\text{sep}}(\{1/2,(K+1)/2\}) = \mathbf{P}_7^{\text{sep}}(\{1,K/2\})$. We conjecture that the $\mathbf{P}_7^{\text{sep}}(\{\tilde\jmath,\tilde\jmath'\})$ is maximal at both $\{\tilde\jmath,\tilde\jmath'\} = \{1/2,(K+1)/2\}$ and $\{1,K/2\}$.}
\end{figure}

For $K=3$, $5$, and $7$, the maximum values of $\mathbf{P}_K^{\text{sep}}$ that appear in Figs.~\ref{fig:heatmap_K3_K5}~and~\ref{fig:heatmap_K7_K9}(a) occur at $\{\tilde\jmath,\tilde\jmath'\} = \{1/2,(K+1)/2\}$. For a closer inspection, line cuts taken along $\tilde\jmath' < 2$ are shown in Figs.~\ref{fig:linecut_K3_K5}~and~\ref{fig:linecut_K7}. We find that for large $\tilde\jmath$, the values of $\mathbf{P}_K^{\text{sep}}$ cluster around a range of values that are strictly smaller than the previously identified maximum score. As such, we can be reasonably sure that the maximum of $\mathbf{P}_K^{\text{sep}}$ over $0\leq\tilde\jmath,\tilde\jmath' \leq \infty$ occurs at $\{\tilde\jmath,\tilde\jmath'\} = \{1/2,(K+1)/2\}$ for $K=3$, $5$, and $7$.


Meanwhile, for $7 \leq K \leq 21$, Figs.~\ref{fig:heatmap_K7_K9}~and~\ref{fig:heatmap_big} clearly show that the maximums occur at $\{\tilde\jmath,\tilde\jmath'\} = \{1,K/2\}$, and the separable bound decreases sharply as $\tilde\jmath$ and $\tilde\jmath'$ increases. We conjecture that this pattern also holds for larger values of $K$, and in conjunction with the previous conjecture, we have the following.
\begin{conjecture}\label{conj:sep-argmax}
The maximum separable bound over $0 < \tilde\jmath,\tilde\jmath' \leq \infty$ for fixed $K$ occurs at
\begin{equation}\label{eq:conj-argmax}
\underset{\{\tilde\jmath,\tilde\jmath'\}}{\operatorname{argmax}}\, \mathbf{P}_K^{\text{\normalfont sep}}(\{\tilde\jmath,\tilde\jmath'\}) =
\begin{cases}
    \Bqty{\frac{1}{2},\frac{K+1}{2}} & \text{for $K \leq 7$,}\\[1ex]
    \Bqty{1,\frac{K}{2}} & \text{for $K \geq 7$.}
\end{cases}
\end{equation}
\end{conjecture}

Having found the probable spins at which the maximum occurs, we shall now attempt to find the maximum separable bounds themselves. For $\{\tilde\jmath,\tilde\jmath'\} = \{1/2,(K+1)/2\}$, the separable bound can be exactly solved to be
\begin{equation}\label{eq:sep-spin-half}
\begin{aligned}
    &\mathbf{P}_K^{\text{sep}}\pqty{\Bqty{\tfrac{1}{2},\tfrac{K+1}{2}}}\\
    &\quad{}={} \begin{cases}
        \frac{23}{32} \qquad \text{if $K=3$, otherwise}\\[1ex]
        \frac{1}{2}\bqty{
            1 + 
            \frac{2^{-(K+1)}}{K+1}\spmqty{K-1\\\frac{K-1}{2}}\pqty{\scriptstyle K + \sqrt{
            3K^2 + 18K + 16
        }}
        }.
    \end{cases}
\end{aligned}
\end{equation}
The derivation of Eq.~\eqref{eq:sep-spin-half} is given in Appendix~\ref{apd:proof-spin-one}. 

\begin{figure}[ht]
    \centering
    \includegraphics{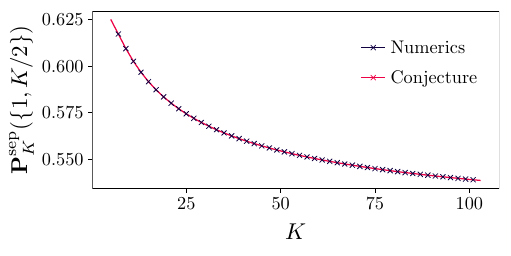}
    \caption{The conjectured and numerically calculated separable bounds for $\{\tilde\jmath,\tilde\jmath'\} = \{1,K/2\}$. The numerical error $\delta \mathbf{P}_K^{\text{sep}} < 10^{-11}$ is too small to be plotted. This is in excellent agreement with Eq.~\eqref{eq:conj-spin-one} for all plotted values of $7 \leq K \leq 101$.}
    \label{fig:spinOne}
\end{figure}

We were unable to apply the same analytical methods to $\{\tilde\jmath,\tilde\jmath'\} = \{1,K/2\}$, but we have a guess for the expression of the separable bound in that case.
\begin{conjecture}\label{conj:sep-spin-one}
    The separable bound for $\{\tilde\jmath,\tilde\jmath'\} = \{1,K/2\}$ with $K \geq 7$ is
    \begin{equation}\label{eq:conj-spin-one}
        \mathbf{P}_K^{\normalfont\text{sep}}\pqty{\Bqty{1,\tfrac{K}{2}}} =
        \frac{1}{2}\bqty{1 + 2^{-(K-1)}\binom{K-1}{\frac{K-1}{2}}\frac{K-1}{K+1}}.
    \end{equation}
\end{conjecture}
Note that $\mathbf{P}_K^{\text{sep}}\pqty{\Bqty{1,\tfrac{K}{2}}} = 1/2$ for $K=3$ and $5$, hence their exclusion from the conjecture. While we were unable to prove Eq.~\eqref{eq:conj-spin-one}, it is in excellent agreement with the numerics up to $K = 101$, as shown in Fig.~\ref{fig:spinOne}. Furthermore, Eqs.~\eqref{eq:sep-spin-half}~and~\eqref{eq:conj-spin-one} are equal for $K=7$, which is consistent with the two maximums in Fig.~\ref{fig:linecut_K7}.

Additionally, by taking the numerical precision into account, Eq.~\eqref{eq:conj-spin-one} can be turned into a reliable upper bound for Fig.~\ref{fig:spinOne}.
\begin{result}\label{res:sep-spin-one}
    The separable bound for $\{\tilde\jmath,\tilde\jmath'\} = \{1,K/2\}$ with $7 \leq K \leq 101$ is upper bounded by $\mathbf{P}_K^{\normalfont\text{sep}}\pqty{\Bqty{1,\tfrac{K}{2}}} + \delta\mathbf{P}_K^{\normalfont\text{sep}}\pqty{\Bqty{1,\tfrac{K}{2}}}$, where $\delta\mathbf{P}_K^{\normalfont\text{sep}} < 10^{-11}$ and
    \begin{equation}\label{eq:conj-spin-one-upper-bounded}
        \mathbf{P}_K^{\normalfont\text{sep}}\pqty{\Bqty{1,\tfrac{K}{2}}} = 
        \frac{1}{2}\bqty{1 + 2^{-(K-1)}\binom{K-1}{\frac{K-1}{2}}\frac{K-1}{K+1}}.
    \end{equation}
\end{result}
We emphasize that Result~\ref{res:sep-spin-one}, while numerical, is a definite result, as the global convergence of semi-definite programs guarantees that the computed values are reliable upper bounds.

\subsubsection{Results and conjectures on separable bounds for general \texorpdfstring{$\sum_{n=1}^N j_n$}{total spin}, fixed \texorpdfstring{$K$}{number of measurements}}
Finally, we can assemble the results and conjectures from the preceding sections to find a separable bound for a spin ensemble $\{j_n\}_{n=1}^N$ with general $\sum_{n=1}^N j_n$ and fixed $K$. Using the upper bounds on the bipartite separable bounds from Results~\ref{res:sep-argmax}~and~\ref{res:sep-spin-one}, an upper bound on the multipartite separable bound can be found with Eq.~\eqref{eq:multipartite-sep-upper-bound}.
\begin{mdframed}
\begin{result}\label{res:bi-K-sep}
Consider an $N$-partite spin ensemble $\{j_n\}_{n=1}^{N}$ such that there are no spin-zero sectors in their bipartitions and $\sum_{n=1}^N j_n \leq 15$. Perform the precession protocol with odd $3 \leq K \leq 21$ on the total angular momentum of the system. If the score $P_K > \mathbf{P}_K^{\text{\normalfont{conj}}} + 10^{-11}$ is obtained, where
\begin{equation*}
\mathbf{P}_K^{\text{\normalfont{conj}}} := \begin{cases}
    \frac{23}{32} & \text{if $K=3$,}\\[1ex]
    \frac{69+\sqrt{181}}{128} &\text{if $K=5$,}\\[1ex]
    \frac{1}{2}\bqty{1 + 2^{-(K-1)}\spmqty{K-1\\\frac{K-1}{2}}\frac{K-1}{K+1}} &\text{otherwise,}
\end{cases}
\end{equation*}
the spin ensemble is GME.
\end{result}
\end{mdframed}
Meanwhile, when the makeup of the spin ensemble is completely unknown, the only possible bound for the total spin is $\sum_{n=1}^N j_n \leq \infty$, and the maximization in Eq.~\eqref{eq:multipartite-sep-upper-bound} would be over all possible $0\leq\tilde\jmath,\tilde\jmath'\leq\infty$. While there are no analytical results in this case, Conjectures~\ref{conj:sep-argmax}~and~\ref{conj:sep-spin-one} can likewise be extended into a conjecture about the multipartite separable bound.
\begin{mdframed}
\begin{conjecture}\label{conj:K-sep}
Consider a spin ensemble with no spin-zero sectors in their bipartitions. Perform the precession protocol with odd $K \geq 3$ on the total angular momentum of the system. If the score $P_K > \mathbf{P}_K^{\text{\normalfont{conj}}}$ is obtained, then the spin ensemble is GME.
\end{conjecture}
\end{mdframed}

\section{Comparison to Other Protocols}

\subsection{Precession protocol as a quantum circuit}
\begin{figure}[ht]
    \centering
    \includegraphics{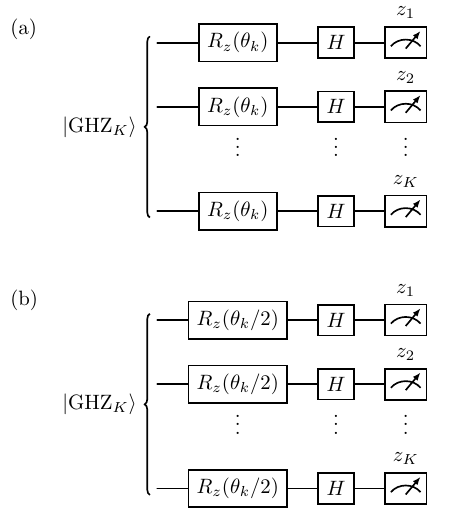}
    \caption{The detection of GHZ entanglement as a quantum circuit, for (a) the precession protocol and (b) the fidelity-based witness. Here, $\theta_k = 2\pi k/K$, $R_z(\theta) = \exp(-i\theta \sigma_z /2)$ is a $z$-rotation gate, $H = (\sigma_x + \sigma_z)/\sqrt{2}$ is the Hadamard gate, and $z_n$ is the outcome of $\sigma_z$ on the $n$th qubit. For both witnesses, a measurement is performed for each $\theta_k$ with $k \in \{1,2,\dots, K \}$, a collective quantity is calculated based on the outcomes $\{z_n\}_{n=1}^K$, and entanglement is detected when the average of this quantity over $k$ violates some separable bound. Our PEW requires $K$ measurement settings while the fidelity-based witness requires $K+1$ settings, and as such the former does not provide a significant improvement over the latter. Therefore, one might be better off using the fidelity-based witness in the context of quantum computation.}
    \label{fig:circuit}
\end{figure}

As the precession protocol can be performed with local spin measurements and classical post-processing, our PEW can also be implemented as a quantum circuit as shown in Fig.~\ref{fig:circuit}(a). In this context, it is interesting to point out the similarities with the fidelity-based witness that is commonly used in the field of quantum computation \cite{GHZ-fidelity}, as is shown in Fig.~\ref{fig:circuit}(b).

(1) Both witnesses involve $K$ equally distributed measurement settings: $\sigma_{\theta_k}$ for our PEW and $\sigma_{\theta_k/2}$ for the fidelity-based witness, where $\sigma_{\theta} = \cos\theta \sigma_x + \sin\theta \sigma_y$, $\sigma_s = J_s^{(j=1/2)}/2\hbar$ are the usual Pauli operators, and $\theta_k = 2\pi k/K$ for $k \in \{1,2,\dots,K\}$.

(2) For each measurement setting $k$, a collective quantity is calculated from $\{z_n\}_{n=1}^K$, where $z_n$ is the measurement outcome of $\sigma_{\theta}$ for the $n$th qubit. This is the majority $\Delta_k \coloneqq \operatorname{maj}(\{z_n\}_{n=1}^K)$ for our PEW, where $\operatorname{maj}(\mathbf{S})$ returns the most commonly occurring value of the set $\mathbf{S}$, and the parity $\tilde{\Delta}_k \coloneqq  (-1)^k\prod_{n=1}^N z_n$ for the fidelity-based witness.

(3) Finally, entanglement is detected in both witnesses when the average of the collective quantity over all measurement settings violates some separable bound. This is $(1+\frac{1}{K}\sum_{k=1}^K\Delta_k)/2 > \mathbf{P}_K^{K\text{-sep}}$ for our PEW, and $(1-\frac{1}{K}\sum_{k=1}^K\tilde{\Delta}_k)/2 < \tilde{\Delta}_0/2$ for the fidelity-based witness, which requires an additional quantity $\tilde{\Delta}_0 \coloneqq \langle (\lvert{\frac{K}{2}, \frac{K}{2}}\rangle\!\langle{\frac{K}{2}, \frac{K}{2}}\rvert  +  \lvert{\frac{K}{2}, -\frac{K}{2}}\rangle\!\langle{\frac{K}{2}, -\frac{K}{2}}\rvert ) \rangle$ that can be computed by measuring all qubits in the $\sigma_z$ basis.

As such, our PEW, which requires measuring in $K$ different local bases, does not offer a significant improvement over the fidelity-based witness, which requires $K+1$ different local bases. Therefore, the fidelity-based witness might be a better choice if only individual addressing is available, as is usual in quantum computation.

\subsection{Nondetection of highly symmetric states}
A $\pi$ rotation around the $z$ axis, which acts as a parity operator for the eigenstates of $J_z$, has the action $e^{-i\pi J_z/\hbar} J_k e^{i\pi J_z/\hbar} = - J_k$ on the $x$--$y$ plane, which implies $e^{-i\pi J_z/\hbar} Q_K e^{i\pi J_z/\hbar} = \mathbbm{1}-Q_K$. Hence, for any state $\rho$ such that $\comm{e^{i\pi J_z/\hbar}}{\rho} = 0$,
\begin{equation}
    \tr(\rho Q_K)
        = \tr( e^{-i\pi J_z/\hbar} \rho e^{i\pi J_z/\hbar} Q_K )
        = 1 - \tr(\rho Q_K).
\end{equation}
As such, $\tr(\rho Q_K) = 1/2 < \mathbf{P}_K^{K\text{-sep}} < \mathbf{P}_K^{\text{conj}}$ for any odd and finite $K$, so $\rho$ does not violate the separable bound.

If $\comm{e^{i\pi J_{\hat{n}}/\hbar}}{\rho} \neq 0$ for another direction $\hat{n}$, the protocol can be performed with measurements in the plane perpendicular to $\hat{n}$, where there might possibly still be a violation of the separable bound.

However, if $\rho$ commutes with the parity operator along every direction, then it will certainly not be detected by our witness. Hence, an important family of states that is missed by our protocol are the Werner states and their multipartite extensions \cite{Werner-states-OG,Werner-states-multi}.

\subsection{Nondetection of Dicke states}
Another well-studied family of entangled states are the Dicke states
\begin{equation}
    \ket{D_l^n} := {\binom{n}{l}}^{-\frac{1}{2}}\sum_{\abs{\mathcal{P}}} \mathcal{P}\pqty{ \ket{\uparrow}^{\otimes l} \otimes \ket{\downarrow}^{\otimes (n-l)} },
\end{equation}
where the sum is over all permutations $\mathcal{P}$ of the $n$ spin-half states, with $\ket{\uparrow} \coloneqq \lvert{\frac{1}{2},\frac{1}{2}}\rangle$ and $\ket{\downarrow} \coloneqq \lvert{\frac{1}{2},-\frac{1}{2}}\rangle$. These include $W$ states as members, and can be detected effectively using just spin angular momentum measurements \cite{korbicz_2006, toth_ssi_2009, vitagliano_sse_2014, toth_dicke_2007, huber_dicke_2011, duan_dicke_2011, lucke_dicke_2014}. Any of the referenced witnesses would be a better choice for detecting the entanglement of such states, as our witness is unable to detect the entanglement of Dicke states in general.

We note an exception: the tripartite Dicke or $W$ state
\begin{equation}
    \ket{W_3} := \ket{D_1^3} = \frac{1}{\sqrt{3}}\pqty{\ket{\uparrow\downarrow\downarrow} + \ket{\downarrow\uparrow\downarrow} + \ket{\downarrow\downarrow\uparrow}}
\end{equation}
achieves $\ev{Q_3}{W_3} = 11/16 > \mathbf{P}_3^{\text{conj}} > \mathbf{P}_3^{3\text{-sep}}$ when performing the precession protocol with measurements in the $y$--$z$ plane. Hence, $\ket{W_3}$ is detected by our PEW, although this does not work beyond the tripartite case.

\subsection{Detection of GHZ states}
It follows directly from Eq.~\eqref{eq:QK-eigen-decomposition} that the precession protocol with $K$ measurements can be used to detect the entanglement of the $K$-partite GHZ state, defined as
\begin{equation}
    \ket{\text{GHZ}_K} := \frac{1}{\sqrt{2}}\pqty{\ket{\uparrow}^{\otimes K} + (-1)^{\frac{K-1}{2}} \ket{\downarrow}^{\otimes K}}.
\end{equation}
In fact, any GHZ state $\propto \ket{\uparrow}^{\otimes K} + e^{i\phi}\ket{\downarrow}^{\otimes K}$ with any phase $e^{i\phi}$ can be detected by the protocol with the replacement $J_k \to e^{-i\theta_i J_z/\hbar}J_ke^{i\theta_i J_z/\hbar}$, where $\theta_i = [(K-1)\pi-2\phi]/(2K)$.

Meanwhile, a GHZ state in the presence of global depolarizing noise is given by 
\begin{equation}
    \rho^{(p_G)}_{\text{GHZ},K} := p_G \frac{\mathbbm{1}}{\tr(\mathbbm{1})} + (1-p_G)\ketbra{\text{GHZ}_K},
\end{equation}
which is GME if and only if $p_G < 1/[2(1-2^{-K})]$ \cite{mixed-GHZ-max-GME}. This state achieves the score
\begin{equation}
    \tr(\rho^{(p_G)}_{\text{GHZ},K} Q_K)
    = \frac{1}{2} + 2(1-p_G) \pqty{\mathbf{P}_{K}^{K\text{-sep}}-\frac{1}{2}}.
\end{equation}
If the total spin of the system is known to be $K/2$, our witness certifies GME when ${\tr}(\rho^{(p_G)}_{\text{GHZ},K} Q_K) > \mathbf{P}_{K}^{K\text{-sep}}$, and hence is detected for the range $p_G < 1/2$. Therefore, our PEW can detect noisy GHZ states close to the theoretical limit.

As mentioned in the Introduction, GHZ states fail to be detected by almost every other angular-momentum-based witness mentioned in this paper \cite{teh_reid_2019,li_gme_2021,toth_singlets_2004,toth_ssi_2007, toth_ssi_2009,vitagliano_sse_2011, vitagliano_sse_2014,toth_spin_2005}. The only exception is $\ket{\text{GHZ}_3}$, whose detection using angular momentum measurements is described in both Refs.~\cite{korbicz_2005}~\&~\cite{teh_reid_2019}; but the reported methods do not work beyond the tripartite case.

\subsection{Other states detected by our PEW}\label{sec:detected}
GHZ states are just one family of several that our PEW can detect, while possessing certain symmetries that prevent their detection by previous angular-momentum-based witness.

For a spin ensemble $\{j_n\}_{n=1}^N$ such that $\sum_{n=1}^N j_n$ is a half integer and $\sum_{n=1}^N j_n > K/2$, the $j=K/2$ subspace will be in general degenerate. Hence, the space of states that are eigenvalues of $Q_K^{(j=K/2)}$ with the same eigenvalue as $\ket{\text{GHZ}_K}$ will have a dimension larger than one, and any state that lives in this subspace will violate the separable bound and hence will be detected. A specific example for a spin-half ensemble with $N=7$ particles that is detected by the precession protocol with $K=5$ is
\begin{equation}
\begin{aligned}
    \ket{\Phi_5} &\coloneqq \frac{1}{2}\sqrt{\frac{7}{3}}\pqty{
\ket{\uparrow}\otimes\ket{\downarrow}^{\otimes 6} -
\ket{\downarrow}\otimes\ket{\uparrow}^{\otimes 6}
} \\
    &\qquad{}-{} \frac{1}{2\sqrt{3}}\pqty{\ket{D_1^7} - \ket{D_6^7}}.
\end{aligned}
\end{equation}
More generally, any state that lives in the subspace spanned by the eigenstates of $Q_K$ whose eigenvalues are larger than the separable bound will be detected by our PEW. An example of this is
\begin{equation}
    \ket{\Phi_3} \coloneqq \frac{\sqrt{3}}{2}\ket{\mathrm{GHZ}_9} - \frac{1}{2\sqrt{2}}\pqty{\ket{D_3^9} + \ket{D_6^9}},
\end{equation}
which is also an eigenstate of $Q_3$ with an eigenvalue that is different from $\bra{\text{GHZ}_3}Q_3\ket{\text{GHZ}_3}$, but which nonetheless satisfies $\bra{\Phi_3}Q_3\ket{\Phi_3} > \mathbf{P}_3^{\text{conj}}$. Hence $\ket{\Phi_3}$, and indeed any superposition of $\ket{\text{GHZ}_3}$ and $\ket{\Phi_3}$, will be detected by the precession protocol with $K=3$. We show in Appendix~\ref{appendix: comparison} that all of the discussed states ($\ket{\text{GHZ}_K}$, $\ket{\Phi_5}$, and $\ket{\Phi_3}$) are missed by all the other previously reported angular-momentum-based criteria, so these states can be detected only by our PEW.

The complete characterization of detectable states is still an open problem. This is partly because the eigendecomposition of each block $Q_K^{(j)}$ is only known analytically for certain values of $j$, and partly because the space of detectable states depends on the degeneracy of each $j$, which, in turn, depends on the exact makeup of the ensemble. Regardless, with the above sufficient conditions, $Q_K$ can be constructed numerically to find the subspace of states that will be detected by our PEW.

\section{Variations on the protocol}

\subsection{Improved separable bound in the presence of more information}
The reported results and conjectures heavily utilizes Eq.~\eqref{eq:multipartite-sep-bound}, which only takes into account the value of $\sum_{n=1}^N j_n$ and the score achieved when performing the precession protocol on the total angular momentum of the spin ensemble. No specific makeup of the spin ensemble $\{j_n\}_{n=1}^N$ was assumed.

Of course, more information can be obtained by characterizing the system in more detail or by performing other measurements. For example, if the system is a collection of bosons, the ensemble will be one of integer spins, and any fixed spin $\tilde\jmath$ that appears in the block decomposition of the total angular momentum must be a positive integer $\tilde\jmath\in \mathbb{Z}_0^+$. Another example would be if the magnitude $\lvert\vec{J}\rvert^2$ of the total angular momentum is measured in addition to performing the precession protocol, and was found take on possible values $j(j+1)$ for some $j \in \{j_{\text{min}},\dots,j_{\text{max}}\}$. Then, the only spins $\tilde\jmath$ and $\tilde\jmath'$ that are compatible with the observed measurements of $\lvert\vec{J}\rvert^2 = \lvert \vec{J}^{(\tilde\jmath)} + \vec{J}^{(\tilde\jmath')} \rvert^2$ are those such that $\lvert\tilde\jmath-\tilde\jmath'\rvert \leq j_{\text{min}}$ and $\tilde\jmath+\tilde\jmath' \geq j_{\text{max}}$.

Just like the above two examples, any additional information about the system can restrict the possible spin pairs that appear in the maximization of Eq.~\eqref{eq:multipartite-sep-bound}. As this results in a maximization over a restricted range of arguments, the separable bound can be possibly lowered, requiring less violation to detect entanglement.

\subsection{Constructing other witnesses using \texorpdfstring{Sec.~\ref{sec:multi-to-bi}}{the relationship between multipartite and bipartite separable bounds}}
A relation between the multipartite and bipartite separable bounds derived in Sec.~\ref{sec:multi-to-bi} was used in the process of constructing our PEW. This relation can be used to construct other entanglement witnesses. For example, consider performing the precession protocol with $K$ measurements on a spin ensemble with total spin $K/2$ for $K$ odd, but with the replacement $\pos(J_k) \to f_0\mathbbm{1} + f_{\text{odd}}(J_k)$ with some odd function $f_{\text{odd}}(-x) = - f_{\text{odd}}(x)$. A similar analysis to Sec.~\ref{sec:sep-bound-K/2} gives the separable bound
\begin{equation}
\begin{aligned}
&\min_{\rho_{\text{sep}}}\tr{\rho_{\text{sep}}\frac{1}{K}\sum_{k=0}^{K-1}\bqty{f_0\mathbbm{1} + f_{\text{odd}}(J_k)}} \\
&\quad{}={} f_0 + \frac{1}{2}\lvert \langle \tfrac{K}{2},\tfrac{K}{2} \rvert f_{\text{odd}}(J_x) \lvert \tfrac{K}{2}, -\tfrac{K}{2} \rangle \rvert
\end{aligned}
\end{equation}
when minimized over separable states $\rho_{\text{sep}}$. This means that GHZ states will be detected by this modified witness as long as $f_K > 0$, which requires $\frac{d^{K+2L}}{dx^{K+2L}}f_{\text{odd}}(x) \rvert_{x=0} \neq 0$ for some integer $L \geq 0$. Of course, by choosing $f_{\text{odd}}(x) \propto 2\pos(x)-1$  as we do here, the magnitudes of the angular momentum measurements do not need to be resolved, which is a benefit that might be lost with other choices of $f_{\text{odd}}(x)$.

On closer inspection, the relation derived in Sec.~\ref{sec:multi-to-bi} is not specific to our PEW, and is in fact applicable to any function $f(\vec{J})$ of angular momentum.

To illustrate this, consider the commonly used witness $\lvert\vec{J}\rvert^2$: given $g^{\text{sep}} := \min_{\rho_{\text{sep}}}{\tr}(\rho_{\text{sep}}\lvert\vec{J}\rvert^2)$ minimized over separable states $\rho_{\text{sep}}$, $\langle\lvert\vec{J}\rvert^2\rangle < g^{\text{sep}}$ implies entanglement \cite{toth_singlets_2004,toth_spin_2005}. This witness corresponds to setting $f(\vec{J}) = -\lvert\vec{J}\rvert^2$ in Eq.~\eqref{eq:max-func-J-multi-vs-bi}, and the bipartite separable bound can be worked out to be
\begin{equation}\label{eq:Jsquared-bipartite}
\begin{aligned}
    & g^{\text{sep}}(\{j_1,j_2\}) \\
    &\quad{}={} \min_{\ket{\psi_{j_1},\psi_{j_2}}}\bra{\psi_{j_1},\psi_{j_2}} \abs{\vec{J}^{(j_1)} + \vec{J}^{(j_2)}}^2\ket{\psi_{j_1},\psi_{j_2}} \\
    &\quad{}={} \min_{\ket{\psi_{j_1},\psi_{j_2}}}\Bigg(
        \overbrace{\bra{\psi_{j_1}}{\abs{\vec{J}^{(j_1)}}^2}\ket{\psi_{j_1}}}^{\hbar^2 (j_1^2 + j_1)} + \overbrace{\bra{\psi_{j_2}}{\abs{\vec{J}^{(j_2)}}^2}\ket{\psi_{j_2}}}^{\hbar^2 (j_2^2 + j_2)} \\
        &\hspace{6em}{}+{} 2\hspace{-0.5em}\underbrace{
        \bra{\psi_{j_1}}\vec{J}^{(j_1)}\ket{\psi_{j_1}} \cdot \bra{\psi_{j_2}}\vec{J}^{(j_2)}\ket{\psi_{j_2}}}_{\begin{aligned}
            &\scriptstyle\geq - \sqrt{ \abs{ \bra{\psi_{j_1}}\vec{J}^{(j_1)}\ket{\psi_{j_1}}}^2
            \abs{\bra{\psi_{j_2}}\vec{J}^{(j_2)}\ket{\psi_{j_2}}}^2 } \\
            &\scriptstyle= - \hbar^2 j_1 j_2 
        \end{aligned}}
    \Bigg)\\
&\quad{}\geq{}\hbar^2 [(j_1-j_2)^2 + (j_1+j_2)],
\end{aligned}
\end{equation}
with equality $g^{\text{sep}}(\{j_1,j_2\}) = \hbar^2[(j_1-j_2)^2 + (j_1+j_2)]$ as setting $\ket{\psi_{j_n}} = \ket{j_n,(-1)^n j_n}$ saturates Eq.~\eqref{eq:Jsquared-bipartite}. Substituting this into Eq.~\eqref{eq:max-func-J-multi-vs-bi} gives $g^{\text{sep}}(\{j_n\}_{n=1}^N) = \hbar^2 \min_{\tilde\jmath,\tilde\jmath'} [(\tilde\jmath-\tilde\jmath')^2 + (\tilde\jmath+\tilde\jmath')]$ for all bipartitions $\mathbf{J}$-$\mathbf{J}^\complement$, $\tilde{\jmath} \in \mathcal{J}(\mathbf{J})$, and $\tilde{\jmath}' \in \mathcal{J}(\mathbf{J}^\complement)$.

As long as the singlet subspace of the chosen spin ensemble is nondegenerate, at least one of $\tilde\jmath$ or $\tilde\jmath'$ will be larger than zero, so $g^{\text{sep}}(\{j_n\}_{n=1}^N) > 0$. Meanwhile, if the singlet subspace exists, the corresponding singlet state achieves $\langle\lvert\vec{J}\rvert^2\rangle = 0 < g^{\text{sep}}(\{j_n\}_{n=1}^N)$, and hence will be detected to be GME. This method is particularly useful for witnessing GME of unequal spins with the observable $\lvert\vec{J}\rvert^2$, which, to the best of our knowledge, has yet to be explored. As an example, take $\{j_n\}_{n=1}^4 = \{\frac{1}{2},\frac{1}{2},1,2\}$. Minimizing over all possible pairs $\{\tilde\jmath,\tilde\jmath'\}$ that arise, we find that $g^{\text{sep}}(\{j_n\}_n)=1$. At the same time, the chosen spins permit a singlet subspace for $j = \lvert(j_1+j_2+j_3)-j_4\rvert = \lvert\frac{1}{2} + \frac{1}{2} + 1 -2 \rvert = 0$, so the GME of the corresponding singlet state can be detected.

More generally, Eq.~\eqref{eq:max-func-J-multi-vs-bi} can be used to construct other GME witnesses by choosing other functions of $f(\vec{J})$. As demonstrated here, this approach allows us to extend existing bipartite witnesses into GME witnesses, which is especially convenient if an analytical form of the bipartite separable bound is known.

\subsection{Bipartite separable bounds When \texorpdfstring{$\tilde\jmath=\tilde\jmath'$}{both spins are the same} for fixed \texorpdfstring{$K$}{number of measurements}}
An interesting observation can be made about the behavior of the separable bound along the $\tilde\jmath=\tilde\jmath'$ diagonal. In Figs.~\ref{fig:heatmap_K3_K5},~\ref{fig:heatmap_K7_K9},~and~\ref{fig:heatmap_big}, the separable bounds along these diagonals are smaller in magnitude than the surrounding values, and seem to stabilize to a limit for large $\tilde\jmath$.

\begin{figure}[ht]
    \centering
    \includegraphics{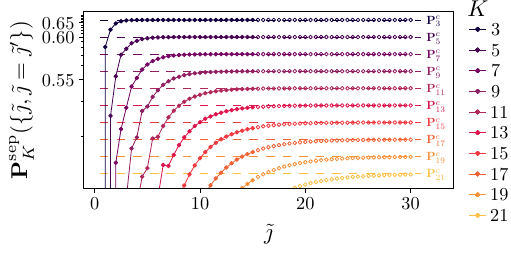}
    \caption{\label{fig:sameSpins}Line cuts along the diagonals $\tilde\jmath = \tilde\jmath'$ along Figs.~\ref{fig:heatmap_K3_K5},~\ref{fig:heatmap_K7_K9},~and~\ref{fig:heatmap_big}. The vertical axis is scaled to fit every case, and the dashed lines are the classical bounds $\mathbf{P}_K^c$. The unfilled points mark those where only the lower bound $\mathbf{P}_K^{\text{sep}} - \delta\mathbf{P}_K^{\text{sep}}$ could be computed. For the filled points, the error bars $\delta\mathbf{P}_K^{\text{sep}}$ are smaller than the marker size. We find that $\mathbf{P}_K^{\text{sep}}(\{\tilde\jmath,\tilde\jmath'\})$ approaches $\mathbf{P}_K^c$ from below as $\tilde\jmath$ increases.}
\end{figure}

Indeed, isolating these diagonals in Fig.~\ref{fig:sameSpins} reveals that $\mathbf{P}_K^{\text{sep}}(\{\tilde\jmath,\tilde\jmath'=\tilde\jmath\})$ approaches the classical bound $\mathbf{P}_K^{c}$ from below as $\tilde\jmath$ increases.

This behavior is in accordance with two previously known results. First, the observable $\pos(J_x^{(j)})$ has the limit $\lim_{j\to\infty}\pos(J_x^{(j)}) = \pos(a + a^\dag)$ for the annihilation operator $a$ of a harmonic oscillator. Second, for two harmonic oscillators with annihilation operators $a_1$ and $a_2$, respectively, if the precession protocol is performed on the normal mode $a_+ \propto a_1 + a_2$, $\mathbf{P}_K^{c}$ is the maximum score achievable by a state separable over the $\{a_1,a_2\}$ modes.

Therefore, as $J_x = J_x^{(\tilde\jmath)} + J_x^{(\tilde\jmath')}$ for $\tilde\jmath = \tilde\jmath'$ appears analogous to $a_+ \propto a_1 + a_2$, it would be expected that $\lim_{\tilde\jmath\to\infty}\pos(J_x) \overset{?}{=} \pos(a_+ + a_+^\dag)$. If so, the separable bound should similarly have the limit $\lim_{\tilde\jmath\to\infty}\mathbf{P}_K^{\text{sep}}(\{\tilde\jmath,\tilde\jmath'=\tilde\jmath\}) \overset{?}{=} \mathbf{P}_K^c$. We have not managed to prove this, but Fig.~\ref{fig:sameSpins} certainly supports this expectation, and suggests the following.
\begin{mdframed}[nobreak=true]
\begin{conjecture}
Consider a bipartite spin ensemble $\{\tilde\jmath,\tilde\jmath'\}$, where $\tilde\jmath=\tilde\jmath'$. Perform the precession protocol with odd $K \geq 3$ on the total angular momentum of the system. If the classical bound $\mathbf{P}_K^c$ is violated, then the two spins are entangled.
\end{conjecture}
\end{mdframed}
While this conjecture is limited to the bipartite witness, it hints at a possible link between the classical bound of the precession protocol, entanglement, and the harmonic oscillator limit.

\section{Conclusion}
In this work, we introduced entanglement witnesses that detect genuine multipartite entanglement of a spin ensemble, which require only measurements of the total angular momentum along several equally spaced directions. We reported analytical expressions for the separable bound when the total spin is a half-integer. For the other cases, reliable numerical values and a conjectured expression is reported for the separable bound, where the latter is well supported by the former.

Of similar angular-momentum-based witnesses, only some detect genuine multipartite entanglement; of those that do, most are effective at detecting Dicke-like states, but none can detect Greenberger-Horne-Zeilinger states beyond the tripartite case. This gap is filled by our PEW, which is effective at detecting precisely those states. We also showed that our witness can detect other GME states, which share similar symmetries to the GHZ state, that are missed by existing criteria.

A possible extension would be to consider the effects of anharmonicities when performing the precession protocol under the assumption of dynamics. A popular experimental platform where spin ensembles naturally arise are spin defects or donors in solid-state materials \cite{spin-defects-review,spin-donors-review}, where quadrupole interactions are present in the Hamiltonian as terms quadratic in angular momentum, which perturbs the dynamics of the system away from a perfectly uniform precession. Such anharmonic effects have been studied for continuous variable systems \cite{zaw2022dynamicsbased}, and many of those findings could be extended to the case of spin angular momentum.

Finally, we once again highlight the generality of the result in Sec.~\ref{sec:multi-to-bi}. By choosing different functions of angular momentum to be used with Eq.~\eqref{eq:max-func-J-multi-vs-bi}, it provides a generic recipe for constructing witnesses of genuine multipartite entanglement out of witnesses of bipartite entanglement. This can be applicable to many other entanglement witnesses completely unrelated to the precession protocol, as we showed with an example in the previous section. This provides another avenue for future study: Since our PEW is mostly effective when the total spin is a half-integer, and cannot witness inseparable states with positive partial transpose, these techniques might allow us to construct a similar witness that will also be effective for integral total spin or bound entangled states.

\section*{Acknowledgments}
Hung Nguyen Quoc is thanked by K.N.H.V.~for his mentorship, and by V.S.~for his hospitality at VNU, Hanoi.

This work is supported by the National Research Foundation, Singapore, and A*STAR under its CQT Bridging Grant. We also thank the National University of Singapore Information Technology for the use of their high performance computing resources.

We thank Haobo Ma and Wenlin Zhang for pointing out a missing assumption in Eq.~\eqref{eq:multipartite-sep-upper-bound} in an earlier version of this work \cite{erratum}.

\bibliography{references}

\appendix
\renewcommand\thefigure{\thesection\arabic{figure}}
\setcounter{figure}{0}

\begin{widetext}
\section{\label{apd:extra-figs}Additional Figures}
Some additional figures are gathered in this appendix. Figure~\ref{fig:heatmap_big} plots $\mathbf{P}_K^{\text{sep}}$ for $11\leq K \leq 21$, which complements Figs.~\ref{fig:heatmap_K3_K5}~and~\ref{fig:heatmap_K7_K9}; Fig.~\ref{fig:heatmap_err} plots the discrepancy $\delta\mathbf{P}_K^{\text{sep}}$ between the numerically computed lower and upper bounds of $\mathbf{P}_K^{\text{sep}}$, as detailed in Appendix.~\ref{appendix: numerical methods}; Fig.~\ref{fig:heatmap_all} plots the lower bound of $\mathbf{P}_K^{\text{sep}}$ for a wider range of $\tilde\jmath$ and $\tilde\jmath'$, demonstrating that the expected behavior of $\mathbf{P}_K^{\text{sep}}$ appears to hold for larger values of $\tilde\jmath$ and $\tilde\jmath'$, providing further support for Conjecture~\ref{conj:K-sep}.
\begin{figure*}[htbp]
    \centering
    \includegraphics{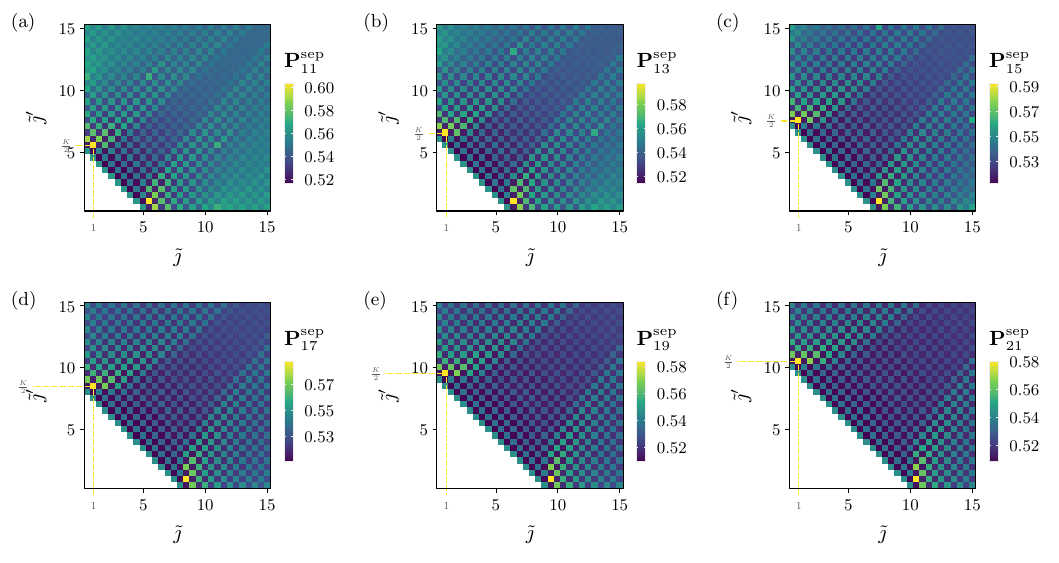}
    \caption{\label{fig:heatmap_big}Heatmap of $\mathbf{P}_{K}^{\text{sep}}(\{\tilde\jmath, \tilde\jmath' \})$ against $\tilde\jmath$ and $\tilde\jmath'$, for (a) $K=11$, (b) $K=13$, (c) $K=15$, (d) $K=17$, (e) $K=19$, and (f) $K=21$. The values $\mathbf{P}_{K}^{\text{sep}}(\{\tilde\jmath, \tilde\jmath' \}) = 1/2$ for $\tilde\jmath + \tilde\jmath' < K/2$ are not plotted here. The separable bounds are large when $\min(\tilde\jmath,\tilde\jmath')$ is small, and $\mathbf{P}_{K}^{\text{sep}}$ decreases as $\min(\tilde\jmath,\tilde\jmath')$ increases. The maximum value of $\mathbf{P}_{K}^{\text{sep}}$ in these cases occur at $\{\tilde\jmath,\tilde\jmath'\} = \{1,K/2\}$.}
\end{figure*}

\begin{figure*}[htbp]
    \centering
    \includegraphics{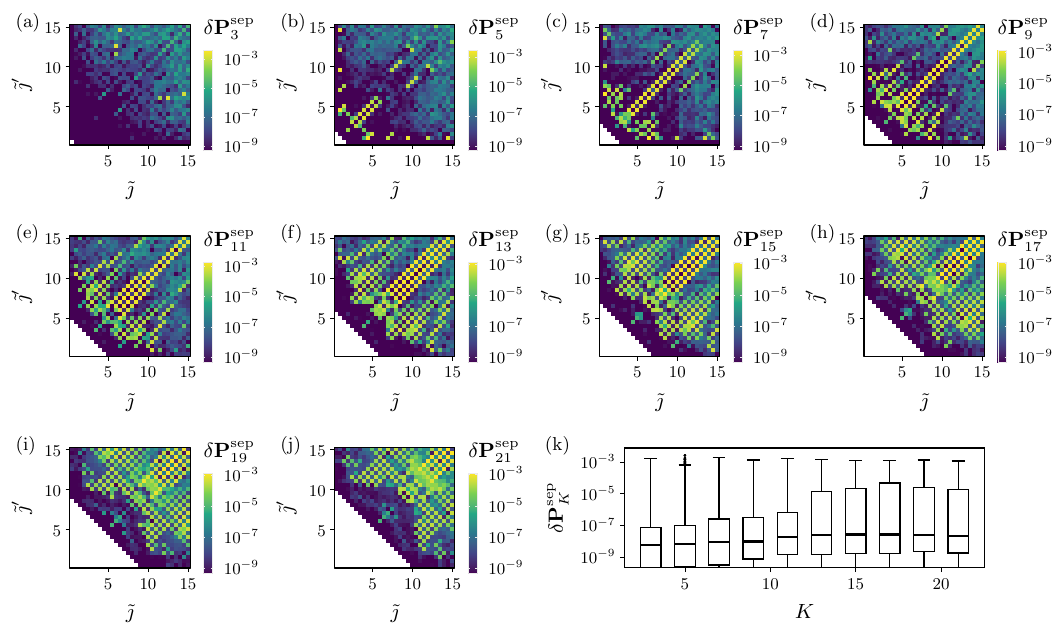}
    \caption{\label{fig:heatmap_err}(a)--(j) Heatmap of $\delta\mathbf{P}_K^{\text{sep}}(\{\tilde\jmath,\tilde\jmath'\})$, the deviations between the lower and upper bounds of $\mathbf{P}_K^{\text{sep}}$ computed using SPI and SDP, respectively, for different values of $K$. The errors are worse for large $j$ as we used larger tolerances in those cases. This is because the value of $\mathbf{P}_K^{\text{sep}}$ is small, so the lower bounds for small $j$ is larger than the upper bound of large $j$ even with the larger tolerances for the latter. (k) Box plots of the errors against the number of measurements. The worst-case errors are of order $10^{-3}$, but most lie within $\delta\mathbf{P}_K^{\text{sep}} \leq 10^{-5}$. Since the upper bound also includes positive-partial-transpose states (see Appendix~\ref{appendix: SDP}), this implies that bound-entangled states cannot be detected by our witness.}
\end{figure*}

\begin{figure*}[htbp]
    \centering
    \includegraphics{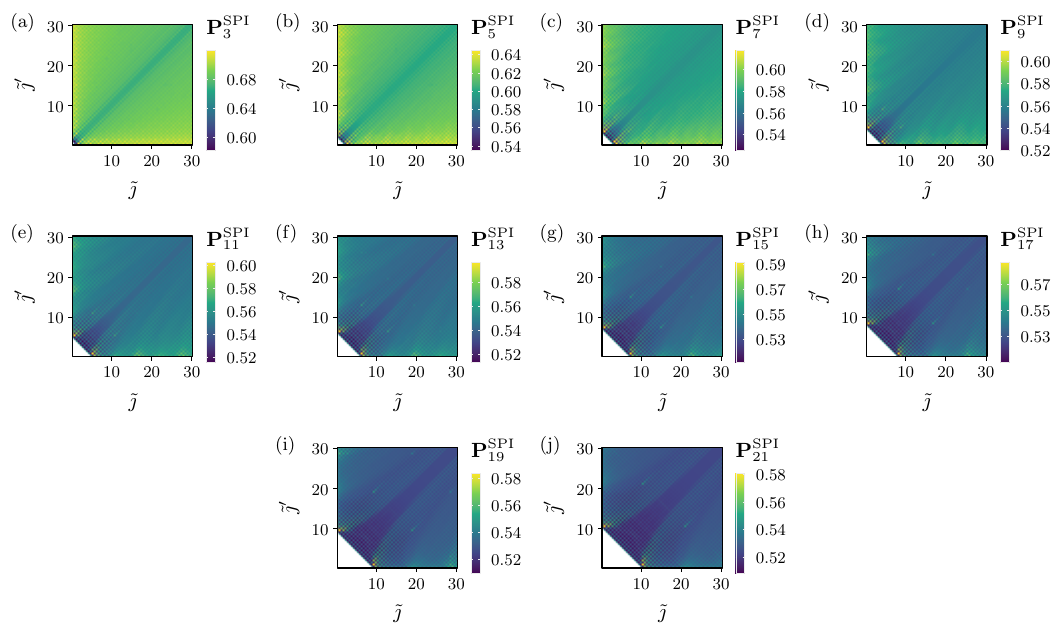}
    \caption{\label{fig:heatmap_all}Heatmap of $\mathbf{P}_K^{\text{SPI}}(\{\tilde\jmath,\tilde\jmath'\}) = \mathbf{P}_K^{\text{sep}}(\{\tilde\jmath,\tilde\jmath'\}) - \delta \mathbf{P}_K^{\text{sep}}(\{\tilde\jmath,\tilde\jmath'\})$, the lower bound of $\mathbf{P}_K^{\text{sep}}$, computed for larger values of $\tilde\jmath$ and $\tilde\jmath'$. We were unable to compute the upper bounds for $\tilde\jmath,\tilde\jmath' > 15$ due to constraints of computational resources. However, from the small deviations between the upper and lower bounds for $\tilde\jmath,\tilde\jmath' \leq 15$ as seen in Fig.~\ref{fig:heatmap_err}, we expect that $\mathbf{P}_K^{\text{SPI}}$ should also be rather close to the actual value $\mathbf{P}_K^{\text{sep}}$. We find that the behavior of $\mathbf{P}_K^{\text{SPI}}$ continues as expected, which is that is reduces in magnitude as $\tilde\jmath$ and $\tilde\jmath'$ increases, which lends further credence to Conjecture~\ref{conj:K-sep}.}
\end{figure*}

\section{Key Technical Aspects of the Precession Protocol}\label{apd:technical-details}
We review some key technical aspects of the precession protocol used in this paper. For the full study with details and derivations, refer to Ref.~\cite{zaw2022dynamicsbased}.

Consider the protocol performed with a single fixed spin $j$ such that the expected score for the state $\rho$ given by $P_K = {\tr}(\rho Q_K^{(j)})$, where
\begin{equation}
    Q_K^{(j)} = \frac{1}{K}\sum_{k=0}^{K-1} \pos(J_k^{(j)})
    = \frac{1}{2}\bqty{
        \mathbbm{1} + \frac{1}{K}\sum_{k=0}^{K-1}
            e^{-i (2\pi k/K) J_z^{(j)}/\hbar} 
            \sgn(J_x^{(j)})
            e^{ i (2\pi k/K) J_z^{(j)}/\hbar}
    }.
\end{equation}
Here, $2\pos(J_k^{(j)})\ket{j,m}_k = \ket{j,m}_k + \sgn(m)\ket{j,m}_k$ and $J_k^{(j)} = e^{-i (2\pi k/K) J_z^{(j)}/\hbar} J_x^{(j)} e^{i  (2\pi k/K) J_z^{(j)}/\hbar}$ were used. Some key properties of $Q_K$ follow.

(1) $Q_K^{(j)}$ is block diagonal. Define the projector
\begin{equation}
    \Pi_K^{(j,\bar{m})} := \sum_{k=0}^{\lfloor({j-\bar{m}})/{K}\rfloor} \ketbra{j,\bar{m}+kK},
\end{equation}
which projects onto the subspace of the eigenstates of $J_z^{(j)}$ whose eigenvalues are $\bar{m}$ modulo $K$. Then, $Q_K^{(j)}$ has the block decomposition
\begin{equation}
Q_K^{(j)} = \frac{1}{2}\bqty{
    \mathbbm{1} + 
    \bigoplus_{\bar{m}=-j}^{\min(j,-j+K-1)} \Pi_K^{(j,\bar{m})} \sgn(J_x^{(j)}) \Pi_K^{(j,\bar{m})}  
}.
\end{equation}
Consequently, if $j<K/2$, then $\bar{m}\leq K/2-1$ and hence $\lfloor({j-\bar{m}})/{K}\rfloor = 0$, which implies that $2Q_K^{(j < K/2)} = \mathbbm{1} + \sum_{m=-j}^j \ketbra{j,m}\sgn(J_x^{(j)})\ketbra{j,m}$ is diagonal with respect to the eigenstates of $J_z$.

(2) The matrix elements $\bra{j,m}\sgn(J_x^{(j)})\ket{j,m'}$ are
\begin{equation}\label{eq:sgn-Jx-mel}
\begin{aligned}
    \mel{j,m}{\sgn(J_x^{(j)})}{j,m'}
    &=
    \begin{cases}
    0 & \text{if $(m-m')\bmod 2 = 0$,} \\
    \frac{(-1)^{\pqty{m'-m-1}/{2}}2^{-(2j-1)}}{m'-m}
    \sqrt{
        \pmqty{
            2\lfloor \frac{j+m}{2} \rfloor \\
             \lfloor \frac{j+m}{2} \rfloor
        }
        \pmqty{
            2\lfloor \frac{j-m}{2} \rfloor \\
             \lfloor \frac{j-m}{2} \rfloor
        }
    } \\
    \qquad\qquad\qquad\qquad{}\times{}\sqrt{
        \pmqty{
            2\lfloor \frac{j+m'}{2} \rfloor \\
             \lfloor \frac{j+m'}{2} \rfloor
        }
        \pmqty{
            2\lfloor \frac{j-m'}{2} \rfloor \\
             \lfloor \frac{j-m'}{2} \rfloor
        }
    } & \text{otherwise.} \\
    \qquad\qquad\qquad\qquad{}\times{}\sqrt{
        (j+m)^{(j+m)\bmod 2}
        (j+m')^{(j+m')\bmod 2}
    } \\
    \qquad\qquad\qquad\qquad{}\times{}\sqrt{
        (j-m)^{(j-m)\bmod 2}
        (j-m')^{(j-m')\bmod 2}
    }
    \end{cases}
\end{aligned}
\end{equation}
In particular, the diagonal elements of $\sgn(J_x^{(j)})$ are zero. From the previous result for $j<K/2$, we therefore have $2Q_K^{(j<K/2)} = \mathbbm{1}^{(j<K/2)}$.

With the above two properties, $Q_K^{(j)}$ can be constructed for any given $j$, and its eigenvectors and eigenvalues can be found using standard numerical techniques. For some cases, this can even be done analytically. A special case we use in this paper is $K/2 \leq j < K$, where
\begin{equation}\label{eq:sgn-Jx-mel-special}
\Pi_K^{(j,\bar{m})}\sgn(J_x^{(j)})\Pi_K^{(j,\bar{m})}
= \begin{cases}
0 & \text{if $j+\bar{m} \leq \lfloor j - K/2 \rfloor$,} \\
\Psi^{(j,\bar{m})}\pqty{
    \lvert{\Psi_+^{(j,\bar{m})}}\rangle\!\langle{\Psi_+^{(j,\bar{m})}}\rvert - 
    \lvert{\Psi_-^{(j,\bar{m})}}\rangle\!\langle{\Psi_-^{(j,\bar{m})}}\rvert
    } & \text{otherwise;}
\end{cases}
\end{equation}
with $\Psi^{(j,\bar{m})} = \lvert\mel{j,\bar{m}}{\sgn(J_x^{(j)})}{j,\bar{m}+K}\rvert$ and $\ket{\Psi_{\pm\bar{m}}} = (\ket{j,\bar{m}} \pm (-1)^{(K-1)/2}\ket{j,\bar{m}+K} )/\sqrt{2}$.

\section{\label{apd:proof-spin-one}Separable Bound for \texorpdfstring{$\{\tilde\jmath,\tilde\jmath'\} = \{1/2,(K+1)/2\}$}{Spin half and (K+1)/2}}
Given two spins with $\{j_1,j_2\} = \{1/2,(K+1)/2\}$, their total angular momentum can take values of $j \in \{ K/2, K/2+1 \}$. Therefore, defining $\Delta Q_K := \mathbbm{1} - 2Q_K = \frac{1}{K}\sum_{k=0}^{K-1}\sgn(J_k)$ for concision,
\begin{align}
    \Delta Q_K &= \bqty{\Pi_K^{\pqty{\frac{K}{2},-\frac{K}{2}}}\operatorname{sgn}(J_x)\Pi_K^{\pqty{\frac{K}{2},-\frac{K}{2}}}} 
    \oplus
    \bqty{\Pi_K^{\pqty{\frac{K}{2}+1,-\frac{K}{2}}}\operatorname{sgn}(J_x)\Pi_K^{\pqty{\frac{K}{2}+1,-\frac{K}{2}}}}
    \\
    &\qquad\qquad{}\oplus{} 
    \bqty{\Pi_K^{\pqty{\frac{K}{2}+1,-\frac{K}{2}-1}}\operatorname{sgn}(J_x)\Pi_K^{\pqty{\frac{K}{2}+1,-\frac{K}{2}-1}}}
    \oplus \bqty{\Pi_K^{\pqty{\frac{K}{2}+1,-\frac{K}{2}+1}}\operatorname{sgn}(J_x)\Pi_K^{\pqty{\frac{K}{2}+1,-\frac{K}{2}+1}}}.\notag
\end{align}
With $\Pi_K^{(j,\bar{m})}\sgn(J_x)\Pi_K^{(j,\bar{m})} =
\Psi^{(j,\bar{m})}(
\lvert{\Psi_+^{(j,\bar{m})}}\rangle\!\langle{\Psi_+^{(j,\bar{m})}}\rvert - 
\lvert{\Psi_-^{(j,\bar{m})}}\rangle\!\langle{\Psi_-^{(j,\bar{m})}}\rvert
)$ from Eq.~\eqref{eq:sgn-Jx-mel-special} and direct computation of $\Psi^{(j,\bar{m})}$ from Eq.~\eqref{eq:sgn-Jx-mel}, we obtain 
\begin{equation}
\begin{aligned}
    \Psi^{(\frac{K}{2},-\frac{K}{2})} &= 2^{-(K-1)}\pmqty{K-1\\\frac{K-1}{2}}, & 
    \Psi^{(\frac{K}{2}+1,-\frac{K}{2})} &= \frac{\Psi^{(\frac{K}{2},-\frac{K}{2})}}{K+1}, &
    \Psi^{(\frac{K}{2}+1,-\frac{K}{2}\mp 1)} &= \sqrt{\frac{K+2}{2(K+1)}}\Psi^{(\frac{K}{2},-\frac{K}{2})}.
\end{aligned}
\end{equation}
We are interested in finding the separable bound
\begin{equation}\label{eq:sep-bound-two-step}
\begin{aligned}
\mathbf{P}_K^{\text{sep}}({\{j_1,j_2}\}) &= \hspace{-1em}\max_{\ket{\psi_{j_1}},\ket{\psi_{j_2}}}(\bra{\psi_{j_1}}\otimes\bra{\psi_{j_2}}) Q_K (\ket{\psi_{j_1}}\otimes\ket{\psi_{j_2}}) \\
&= \frac{1}{2} + \frac{1}{2}\max_{\ket{\psi_{j_1}}}\Bqty{\max_{\ket{\psi_{j_2}}} \bra{\psi_{j_2}}\bqty{\pqty{\bra{\psi_{j_1}}\otimes\mathbbm{1}^{(j_2)}}\Delta Q_K\pqty{\ket{\psi_{j_1}}\otimes\mathbbm{1}^{(j_2)}}} \ket{\psi_{j_2}}}.
\end{aligned}
\end{equation}
By rewriting it as above, the maximization can be thought of as a two-step process: we first find the maximum eigenvalue of
$(\bra{\psi_{j_1}}\otimes\mathbbm{1}^{(j_2)})\Delta Q_K(\ket{\psi_{j_1}}\otimes\mathbbm{1}^{(j_2)})$ for a fixed $\ket{\psi_{j_1}}$, then maximize this eigenvalue over all states $\ket{\psi_{j_1}}$. Parameterizing a generic state in $\mathcal{H}^{(j_1)}$ as $\ket{\psi_{j_1}}:=\cos(\frac{\theta}{2})\ket{\frac{1}{2},-\frac{1}{2}} + e^{i\phi}\sin(\frac{\theta}{2})\ket{\frac{1}{2},\frac{1}{2}}$, and using the Clebsch--Gordan coefficients, we find
\begin{subequations}
\begin{align}
\pqty{\bra{\psi_{j_1}}\otimes\mathbbm{1}^{(j_2)}}\lvert{\Psi^{(\frac{K}{2},-\frac{K}{2})}_{\pm}}\rangle &= \cos(\frac{\theta}{2})\sqrt{\frac{K+1}{2(K+2)}}\ket{\tfrac{K+1}{2},-\tfrac{K+1}{2}} - \frac{e^{-i\phi}\sin(\frac{\theta}{2})}{\sqrt{2(K+2)}}\ket{\tfrac{K+1}{2},-\tfrac{K-1}{2}} \\
&\quad\qquad{}\pm{}(-1)^{\frac{K-1}{2}}\pqty{
    \frac{\cos(\frac{\theta}{2})}{\sqrt{2(K+2)}}\ket{\tfrac{K+1}{2},\tfrac{K-1}{2}} 
    - e^{-i\phi}\sin(\frac{\theta}{2})\sqrt{\frac{K+1}{2(K+2)}}\ket{\tfrac{K+1}{2},\tfrac{K+1}{2}}
},\notag\\
\pqty{\bra{\psi_{j_1}}\otimes\mathbbm{1}^{(j_2)}}\lvert{\Psi^{(\frac{K}{2}+1,-\frac{K}{2})}_{\pm}}\rangle &= \frac{\cos(\frac{\theta}{2})}{\sqrt{2(K+2)}}\ket{\tfrac{K+1}{2},\tfrac{K+1}{2}} 
+ e^{-i\phi}\sin(\frac{\theta}{2})\sqrt{\frac{K+1}{2(K+2)}}\ket{\tfrac{K+1}{2},\tfrac{K-1}{2}} \\
&\quad\qquad{}\pm{}(-1)^{\frac{K-1}{2}}\pqty{\cos(\frac{\theta}{2})\sqrt{\frac{K+1}{2(K+2)}}\ket{\tfrac{K+1}{2},-\tfrac{K-1}{2}} +
\frac{e^{-i\phi}\sin(\frac{\theta}{2})}{\sqrt{2(K+2)}}\ket{\tfrac{K+1}{2},-\tfrac{K+1}{2}} }, \notag\\
\pqty{\bra{\psi_{j_1}}\otimes\mathbbm{1}^{(j_2)}}\lvert{\Psi^{(\frac{K}{2}+1,-\frac{K}{2}-1)}_{\pm}}\rangle &= \cos(\frac{\theta}{2})\sqrt{\frac{K}{2(K+2)}}\ket{\tfrac{K+1}{2},\tfrac{K-3}{2}} + \frac{e^{-i\phi}\sin(\frac{\theta}{2})}{\sqrt{K+2}}\ket{\tfrac{K+1}{2},\tfrac{K-1}{2}} \\
&\quad\qquad{}\pm{} (-1)^{\frac{K-1}{2}}\frac{e^{-i\phi}\sin(\frac{\theta}{2})}{\sqrt{2}}\ket{\tfrac{K+1}{2},-\tfrac{K+1}{2}}, \notag\\
\pqty{\bra{\psi_{j_1}}\otimes\mathbbm{1}^{(j_2)}}\lvert{\Psi^{(\frac{K}{2}+1,-\frac{K}{2}+1)}_{\pm}}\rangle &= \frac{\cos(\frac{\theta}{2})}{\sqrt{K+2}}\ket{\tfrac{K+1}{2},-\tfrac{K-1}{2}}
+ e^{-i\phi}\sin(\frac{\theta}{2}) \sqrt{\frac{K}{2(K+2)}}\ket{\tfrac{K+1}{2},-\tfrac{K-3}{2}} \\
&\quad\qquad{}\pm{} (-1)^{\frac{K-1}{2}}\frac{\cos(\frac{\theta}{2})}{\sqrt{2}}\ket{\tfrac{K+1}{2},\tfrac{K+1}{2}}.\notag
\end{align}
\end{subequations}
Notice therefore that $(\bra{\psi_{j_1}}\otimes\mathbbm{1}^{(j_2)})\Delta Q_K(\ket{\psi_{j_1}}\otimes\mathbbm{1}^{(j_2)})$ has support only in the subspace spanned by $\{\ket{\tfrac{K+1}{2},-\tfrac{K+1}{2}}$, $\ket{\tfrac{K+1}{2},-\tfrac{K-1}{2}}$, $\ket{\tfrac{K+1}{2},-\tfrac{K-3}{2}}$, $\ket{\tfrac{K+1}{2},\tfrac{K-3}{2}}$, $\ket{\tfrac{K+1}{2},\tfrac{K-1}{2}}$, $\ket{\tfrac{K+1}{2},\tfrac{K+1}{2}}\}$. For $K>3$, in this six-dimensional subspace,
\begin{equation}
    \frac{(\bra{\psi_{j_1}}\otimes\mathbbm{1}^{(j_2)})\Delta Q_K(\ket{\psi_{j_1}}\otimes\mathbbm{1}^{(j_2)})}{\Psi^{(\frac{K}{2},-\frac{K}{2})}} \widehat{=}
    \sqrt{\frac{K/2}{K+1}}
    \left(\begin{matrix}
    0 & 0 & 0 & e^{- i \phi} \sin(\theta) & 1 & - \sqrt{\frac{K/2}{K + 1}} e^{i \phi} \sin(\theta) \\
    0 & 0 & 0 & 0 & 0 & 1\\
    0 & 0 & 0 & 0 & 0 & e^{- i \phi} \sin(\theta) \\
    e^{i \phi} \sin(\theta) & 0 & 0 & 0 & 0 & 0\\
    1 & 0 & 0 & 0 & 0 & 0\\
    - \sqrt{\frac{K/2}{K + 1}} e^{i \phi} \sin(\theta) & 1 & e^{i \phi} \sin(\theta) & 0 & 0 & 0\end{matrix}\right).
\end{equation}
This matrix is highly sparse, and its eigenvalues can be found by directly solving the characteristic equation $\det[(\bra{\psi_{j_1}}\otimes\mathbbm{1}^{(j_2)})\Delta Q_K(\ket{\psi_{j_1}}\otimes\mathbbm{1}^{(j_2)})-\lambda\mathbbm{1}] = 0$, which is given by
\begin{equation}
    \pqty{\frac{\lambda}{\Psi^{(\frac{K}{2},-\frac{K}{2})}}}^2\Bqty{
        \pqty{\frac{\lambda}{\Psi^{(\frac{K}{2},-\frac{K}{2})}}}^4 (K+1)^2
        - \pqty{\frac{\lambda}{\Psi^{(\frac{K}{2},-\frac{K}{2})}}}^2\bqty{ \frac{2K+1}{4}K\sin^2(\theta) + 2(K + 1) }
        + \pqty{\frac{K\sin^2(\theta)}{8} + 1}^2
    } = 0.
\end{equation}
Removing the $\lambda=0$ solutions, this equation is quadratic in $\lambda^2$, with the solutions
\begin{equation}
    \pqty{\frac{\lambda(\theta)}{\Psi^{(\frac{K}{2},-\frac{K}{2})}}}^2 =  \pqty{\frac{K\sin(\theta)\pm\sqrt{
        (3K+1)K\sin^2(\theta) + 16(K+1)
    }}{4(K+1)}}^2.
\end{equation}
Among the four solutions for $\lambda(\theta)$, whose dependence on $\ket{\psi_{j_1}}$ that is present through $\theta$ is now made explicit, we clearly have
\begin{equation}
    \frac{\lambda_{\text{max}}(\theta)}{\Psi^{(\frac{K}{2},-\frac{K}{2})}} = \frac{K\abs{\sin(\theta)} + \sqrt{
        (3K+1)K\sin^2(\theta) + 16(K+1)
    }}{4(K+1)}.
\end{equation}
To maximize this over $\theta$, we only need to consider the range $\theta \in [0,\pi/2]$, as $\lambda_{\text{max}}(\theta)$ repeats outside this range. Furthermore, since $\lambda_{\text{max}}(\theta)$ is monotonically increasing with $\theta$ in this range, we arrive at
\begin{equation}
    \max_\theta \lambda_{\text{max}}(\theta) = \frac{2^{-(K+1)}}{K+1}\pmqty{K-1\\\frac{K-1}{2}}\pqty{K + \sqrt{
        3K^2 + 18K + 16
    }}.
\end{equation}
Note that the above is true only for $K>3$. When $K=3$, $\tfrac{K-3}{2} = -\tfrac{K-3}{2} = 0$, so the subspace is only five-dimensional. Nonetheless, repeating the above steps lead to a similar characteristic equation quadratic in $\lambda^2$, with
\begin{equation}
    \max_\theta\lambda_{\text{max}}(\theta) = \max_\theta\frac{3\abs{\sin(\theta)} + \sqrt{
        57\sin^2(\theta) + 64
    }}{32} = \frac{7}{16}.
\end{equation}
As such, recalling that $\max_{\theta}\lambda_{\text{max}}(\theta)$ is the separability bound for $\Delta Q_K$ and $Q_K = (\mathbbm{1} + \Delta Q_K)/2$, we arrive at
\begin{equation}
\mathbf{P}_K^{\text{sep}}\pqty{\Bqty{\tfrac{1}{2},\tfrac{K+1}{2}}} = \frac{1}{2}( 1 + \lambda_K), \quad\text{where}\quad\lambda_K =\begin{cases}
\dfrac{7}{16} & \text{if $K = 3$,}\\[3ex]
\displaystyle \frac{2^{-(K+1)}}{K+1}\pmqty{K-1\\
\frac{K-1}{2}}\pqty{K + \sqrt{
            3K^2 + 18K + 16
        }} & \text{otherwise.}
\end{cases}
\end{equation}
\end{widetext}

\section{Numerical Methods}\label{appendix: numerical methods}
Two numerical methods were used to find the lower and upper bounds of $\mathbf{P}^{\text{sep}}_K(\{j_1,j_2\})$ for general $j_1, j_2$, which allows us to obtain reliable values of the separable bound to within numerical precision. We use a variant of the separability power iteration (SPI) \cite{spi} to obtain a lower bound $\mathbf{P}_K^{\text{SPI}}$, and semidefinite programming (SDP) \cite{boyd_1996} to obtain an upper bound $\mathbf{P}_K^{\text{SDP}}$. Note that the methods, while numerical, return reliable lower and upper bounds, respectively. The scripts used and generated data are available in Ref.~\cite{scripts}, while the implementation details of these two numerical techniques will be covered in the following sections.

After obtaining the two values, we can further define $\mathbf{P}_K^{\text{sep}} := (\mathbf{P}_K^{\text{SDP}}+\mathbf{P}_K^{\text{SPI}})/2$ and $\delta\mathbf{P}_K^{\text{sep}} := (\mathbf{P}_K^{\text{SDP}}-\mathbf{P}_K^{\text{SPI}})/2$. It is guaranteed that the actual value of the separability bound lies within the range $[\mathbf{P}_K^{\text{sep}} - \delta\mathbf{P}_K^{\text{sep}}, \mathbf{P}_K^{\text{sep}} + \delta\mathbf{P}_K^{\text{sep}}]$: henceforth, we will treat $\mathbf{P}_K^{\text{sep}}$ as an estimate for the true value of the separable bound, while $\delta\mathbf{P}_K^{\text{sep}}$ will be treated as a numerical error due to numerical precision and possible gaps between the two bounds.

For all studied cases, the gaps between the results obtained through these two numerical methods are small, where the largest deviation is of the order $\delta\mathbf{P}_K^{\text{sep}}\sim 10^{-3}$. These deviations are plotted for odd $K \leq 21$ in Fig.~\ref{fig:heatmap_err}.

\subsection{Lower Bounds: Separability Power Iteration}
The lower bound can be found by noticing that $\mathbf{P}^{\text{sep}}_K(\{j_1,j_2\})$ is the global maximum of the product numerical range $\{\langle \psi_{j_1}, \psi_{j_2} \rvert Q_K \lvert \psi_{j_1}, \psi_{j_2} \rangle : \lvert \psi_{j_1}, \psi_{j_2} \rangle = \lvert \psi_{j_1}\rangle \otimes \lvert\psi_{j_2} \rangle,\lvert \psi_{j}\rangle \in \mathcal{H}^{(j)}\}$. As such, any particular choice of $\lvert \psi_{j_1}, \psi_{j_2} \rangle$ must necessarily give a lower bound $\langle \psi_{j_1}, \psi_{j_2} \rvert Q_K \lvert \psi_{j_1}, \psi_{j_2} \rangle \leq \mathbf{P}^{\text{sep}}_K(\{j_1,j_2\})$.

By maximizing $\langle \psi_{j_1}, \psi_{j_2} \rvert Q_K \lvert \psi_{j_1}, \psi_{j_2} \rangle$ over $\lvert \psi_{j_1}, \psi_{j_2} \rangle$ using any method, even if only a local maximum is reached, we can still obtain a lower bound that is as large as possible. As mentioned, we use a variant of the separability power iteration (SPI) \cite{spi} to obtain the lower bound $\mathbf{P}_K^{\text{SPI}}$. We detail the technique in Algorithm \ref{alg: SPI}. 
\begin{algorithm}[H]
\caption{Bipartite variant of SPI}
\label{alg: SPI}
\begin{algorithmic}
  \Require{$\ket{a_0} \in \mathcal{H}^{(j_1)}$, $0 < \varepsilon$}
  \State $p_{\text{prev}} \gets -\infty$
  \State $p_{\text{curr}} \gets \infty$
  \State $\ket{a} \gets \ket{a_0}$
  \State $\ket{b} \gets \ket{b_0} \in \mathcal{H}^{(j_2)}$ \hspace{1em}\Comment{Any $\ket{b_0} \in \mathcal{H}^{(j_2)}$ will do}
 \While{$|p_{\text{curr}}-p_{\text{prev}}| \geq \varepsilon$}
  \State $\ket{b} \gets \operatorname{argmax}_{\tilde{\ket{b}} \in \mathcal{H}^{(j_2)}} \tilde{\bra{b}}\operatorname{tr}_{j_1}\!\!\left[  Q_K \left(\ketbra{a} \otimes \mathbbm{1}^{(j_2)} \right)\right] \tilde{\ket{b}}$ 
  \State $\ket{a} \gets \operatorname{argmax}_{\ket{\tilde{a}} \in \mathcal{H}^{(j_1)}} \bra{\tilde{a}}\operatorname{tr}_{j_2}\!\!\left[ Q_K \left(\mathbbm{1}^{(j_1)} \otimes \ketbra{b}\right)\right] \ket{\tilde{a}}$
  \State $p_{\text{prev}} \gets p_{\text{curr}}$
  \State $p_{\text{curr}} \gets \bra{a,b}Q_K\ket{a,b}$
 \EndWhile
    \State \Return{$p_{\text{curr}}$}
\end{algorithmic}
\end{algorithm}
Here $\tr_{j_n}$ is the partial trace over $\mathcal{H}^{(j_n)}$, while the steps requiring $\operatorname{argmax}$ are eigenvalue problems that can be solved with standard numerical libraries.

Algorithm \ref{alg: SPI} halts when the convergence condition $\lvert p_{\text{curr}}-p_{\text{prev}} \rvert < \varepsilon$ for a target precision $\varepsilon$ is met. It can be shown that the algorithm is always convergent, and returns a local maximum for the given starting vector $\ket{a_0} \in \mathcal{H}^{(j_1)}$. Denote the vectors obtained in the $n$th round of Algorithm \ref{alg: SPI} by $\ket{a_i}$ and $\ket{b_i}$, and let $p_i = \bra{a_i, b_i} Q_K \ket{a_i, b_i}$. Then,
\begin{align}
    p_{i+1} &= \bra{a_{i+1}, b_{i+1}} Q_K \ket{a_{i+1}, b_{i+1}} \nonumber\\
    &= \bra{a_{i+1}}\tr_{j_2}\!\!\bqty{
        Q_K \pqty{
            \mathbbm{1}^{(j_1)} \otimes \ketbra{b_{i+1}}
        }
    }\ket{a_{i+1}} \nonumber\\
    &\geq \bra{a_{i}}\tr_{j_2}\!\!\bqty{
        Q_K \pqty{
            \mathbbm{1}^{(j_1)} \otimes \ketbra{b_{i+1}}
        }
    }\ket{a_{i}} \\
    &= \bra{b_{i+1}}\tr_{j_1}\!\!\bqty{
        Q_K \pqty{
            \ketbra{a_i} \otimes \mathbbm{1}^{(j_2)}
        }
    }\ket{ b_{i+1}} \nonumber\\
    &\geq \bra{a_{i}, b_{i}} Q_K \ket{a_{i}, b_{i}}
    = p_i \nonumber
\end{align}
This proves that $\{p_{i}\}_{i \geq 0}$ is monotonically increasing. Moreover, since $Q_K$ is an observable that describes a probability and is therefore bounded from above, the sequence $\{p_{i}\}_{i\geq 0}$, which is a sequence of expectations of $Q_K$, is also bounded. Therefore, the monotonic sequence $\{p_{i}\}_{i\geq0}$ converges, and hence Algorithm~\ref{alg: SPI} will eventually halt for any starting vector $\ket{a_0} \in \mathcal{H}^{(j_1)}$.

Since the local maximum $p$ we obtain depends on our choice of $\ket{a_0}$, we perform the algorithm with many different starting vectors $\{\ket{a_0},\ket{a_0'},\ket{a_0''},\dots\}$ to obtain a set of local maximums $\{p,p',p'',\dots\}$. Following Ref.~\cite{spi}, we chose $\{\ket{a_0},\ket{a_0'},\ket{a_0''},\dots\}$ to be the eigenvectors of generalized Gell-Mann matrices of dimension $2j_1+1$, which spans the full operator space in $\mathcal{H}^{(j_1)}$. Finally, we define $\mathbf{P}_K^\text{SPI} := \max\{p,p',p'',\dots\}$, which provides us with the best lower bound $\mathbf{P}_K^\text{SPI} \leq \mathbf{P}_K^{\text{sep}}$ that was found from the many runs of the algorithm.

\subsection{Upper Bounds: Semi-Definite Programming}\label{appendix: SDP}
The upper bound is due to the Peres--Horodecki criterion, which states that the set of separable states $\{\rho_{\text{SEP}} : \rho_{\text{SEP}} = \sum_{k} p_k \rho_{j_1,k} \otimes \rho_{j_2,k}\}$ is a subset of the set of positive-partial-transpose states $\{\rho_{\text{PPT}} : \rho_{\text{PPT}}^{\Gamma_2} \succeq 0\}$ \cite{peres_separability_1996, horodecki_separability_1996}. Here,  $\rho_{\text{PPT}}^{\Gamma_2}$ is the partial transposition of $\rho_{\text{PPT}}$ whose matrix elements satisfy
\begin{equation}\bra{\psi_{j_1}, \psi_{j_2}}\rho_{\text{PPT}}^{\Gamma_2}\ket{\phi_{j_1}, \phi_{j_2}} = \bra{\psi_{j_1}, \phi_{j_2}}\rho_{\text{PPT}}\ket{\phi_{j_1}, \psi_{j_2}}.\end{equation}
As $\mathbf{P}_K^{\text{sep}}$ involves a maximization over a subset of the positive-partial-tranpose states, $\max_{\rho_{\text{PTT}}}\tr(\rho_{\text{PTT}}Q_K)$ is an upper bound for $\mathbf{P}_K^{\text{sep}}$. Note that this would be a loose upper bound in general: there exists bound entangled states with positive partial transpose which are nonetheless inseparable \cite{bound-entanglement,bound-entanglement-GME}. In turn, a reliable upper bound on $\max_{\rho_{\text{PTT}}}\tr(\rho_{\text{PTT}}Q_K)$ can be obtained using semidefinite programming (SDP) \cite{boyd_1996}, which we shall denote as $\mathbf{P}_K^{\text{SDP}}$. Being an upper bound of the upper bound for $\mathbf{P}_K^{\text{sep}}$, it is of course the case that $\mathbf{P}_K^{\text{SDP}} \geq \mathbf{P}_K^{\text{sep}}$.

Semidefinite programs are linear optimization problems of the form 
\begin{equation}\label{eq: SDP formulation}
\begin{array}{rl}
\displaystyle\max_{X\succeq 0} & \displaystyle\tr(CX) \\
\displaystyle\text{subject to} & \displaystyle\tr(A_lX) = b_l \text{ for $l=1,2,\dots$,} \\
& \displaystyle X \succeq 0,
\end{array}
\end{equation}
where $C$ and $A_l$ are Hermitian operators. We refer to the above as the primal problem, and there is a related dual problem given by
\begin{equation}
\begin{array}{rl}
\displaystyle\min_{y_1,y_2,\dots} & \sum_{l} b_l y_l  \eqqcolon \vec{b}\cdot\vec{y}\\
\text{subject to} & \sum_l y_l A_l \succeq C.
\end{array}
\end{equation}
The weak duality theorem states that the dual objective always upper bounds the primal one: that is, $\vec{b}\cdot\vec{y} \geq \tr(CX)$. Weak duality therefore ensures that the global optimum of both the primal and dual problems are guaranteed to be between the gap given by the two objectives. As such, this is used as a convergence condition when numerically solving SDPs: local optimizations of the primal and dual problems are performed until $|\vec{b}\cdot\vec{y} - \tr(CX)| < \varepsilon$ for a target precision $\varepsilon$.

For our purposes, we are interested in the optimization
\begin{equation}\label{eq: optimization over PPT}
\begin{array}{rl}
            \displaystyle \max_{\rho_{\text{PPT}}} &   \displaystyle  \tr(\rho_{\text{PPT}} Q_K)\\
\text { subject to } & \displaystyle \rho_{\text{PPT}} \succeq  0\\
                       & \displaystyle \rho_{\text{PPT}}^{\Gamma_2} \succeq 0.
\end{array}
\end{equation}
By setting $X = \rho_{\text{PPT}} \oplus \rho_{\text{PPT}}^{\Gamma_2}$, $C = Q_K \oplus 0$, and, for a Hermitian operator basis $\{O_l\}_l$ of $\mathcal{H}^{(j_1)}\otimes\mathcal{H}^{(j_2)}$, $A_l = O_l \oplus (-O_l^{\Gamma_2})$, Eq.~\eqref{eq: optimization over PPT} can be rewritten into the form given in Eq.~\eqref{eq: SDP formulation}. As such, our problem is an SDP, which can be solved by any SDP solver (we used \texttt{JuMP} \cite{JuMP} and \texttt{COSMO} \cite{COSMO} for the reported results). The objective value of the solution to the dual problem gives us a reliable upper bound $\mathbf{P}_K^{\text{SDP}} \geq \max_{\rho_{\text{PPT}}}\tr(\rho_{\text{PPT}} Q_K) \geq \mathbf{P}^{\text{sep}}_K$.

\section{Comparison to Other Criteria}\label{appendix: comparison}
In this appendix, we explicitly show that the states given as examples in Sec.~\ref{sec:detected} cannot be detected by other angular-momentum-based criteria. For reference, these are
\begin{equation}
\label{eq: detected states}
\begin{aligned}
    \ket{\text{GHZ}_K} &= \frac{1}{\sqrt{2}}\pqty{
        \bigotimes_{n=1}^N \ket{j_n,j_n}
    + (-1)^{\frac{K-1}{2}}\bigotimes_{n=1}^N \ket{j_n, - j_n}
    }, \\
    \ket{\Phi_5} &= \frac{1}{2}\sqrt{\frac{7}{3}}\pqty{
    \ket{\uparrow}\otimes\ket{\downarrow}^{\otimes 6} -
    \ket{\downarrow}\otimes\ket{\uparrow}^{\otimes 6}
    } \\
    &\qquad{}-{}\frac{1}{2\sqrt{3}}\pqty{\ket{D_1^7} - \ket{D_6^7}}, \\
    \ket{\Phi_3} &= \frac{\sqrt{3}}{2}\ket{\mathrm{GHZ}_9} - \frac{1}{2\sqrt{2}}\pqty{\ket{D_3^9} + \ket{D_6^9}}.
\end{aligned}
\end{equation}
The subscript in the label denotes the number of measurements $K$ with which to perform the precession protocol.

It can be verified that these example states are, by construction, eigenstates of $Q_K^{(j)}$, where $Q_K = \bigoplus_j Q_K^{(j)}$ and $Q_K^{(j)}$ acts on an irreducible subspace of $\lvert \vec{J} \rvert^2$.

In the rest of this section, we will use $\ket{\Psi_K} \in \{ \ket{\text{GHZ}_K}, \ket{\Phi_3}, \ket{\Phi_5} \}$ to discuss properties common to all three states. As we will be discussing them in some generality, these discussions will also hold true for similar states with the same properties, for example, other simultaneous eigenstates of $Q_K$ and $\lvert \vec{J} \rvert^2$.

\subsection{Some Properties of These Example States}\label{apd:symmetry-properties}
To derive some general properties of $\ket{\Psi_K}$, we highlight two symmetries of $Q_K$: $e^{-i 2\pi J_z/\hbar K} Q_K e^{i2\pi J_z/\hbar K} = Q_K$ and $e^{-i\pi J_x/\hbar} Q_K e^{i\pi J_x/\hbar} = Q_K$.

These can be verified with direct calculation, but also seen more intuitively: the chosen rotations applied on $Q_K$ merely reshuffles, up to the modulus of $T$, the times $\{t_k\}_{k=0}^{K-1}$ over which the average of $\pos(J_k)$ is calculated, which leaves $Q_K$ invariant.

\subsubsection{All example states have \texorpdfstring{$\ev{\sum_{j \in \mathbf{J}}\vec{J}^{(j)}} = 0$}{zero average angular momentum}}\label{apd:example-states-zero-J}
From the previous discussion, we know that $Q_K$ commutes with both $e^{i\pi J_x/\hbar}$, and therefore the two observables can be simultaneously diagonalized. Indeed, the given example states can be verified to also be eigenstates of $e^{-i\pi J_x/\hbar}$ with eigenvalues $\pm (-1)^j$.

This leads us to
\begin{equation}
\begin{aligned}
    \bra{\Psi_K}J_z^{(j_n)}\ket{\Psi_K}
    &= \bra{\Psi_K}e^{-i\pi J_x/\hbar}J_z^{(j_n)}e^{i\pi J_x/\hbar}\ket{\Psi_K} \\
    &= -\bra{\Psi_K}J_z^{(j_n)}\ket{\Psi_K},
\end{aligned}
\end{equation}
which implies $\langle{J_z^{(j_n)}}\rangle = 0$, and hence $\langle{\sum_{j \in \mathbf{J}}J_z^{(j)}}\rangle = 0$. Replacing $J_z \to J_y$ also gives $\langle{\sum_{j \in \mathbf{J}}J_y^{(j)}}\rangle = 0$.

Meanwhile, the commutation and hence simultaneous diagonalizability of $Q_K$ and $e^{i2\pi J_z/\hbar K}$ similarly lets us choose the example states to be eigenstates of $e^{-i 2\pi J_z/K}$ with eigenvalue $e^{-i 2\pi m/K}$ for some $m$. Then, for any integer $k$,
\begin{equation}
\begin{aligned}
    &\bra{\Psi_K}{
        J_x^{(j_n)}
    }\ket{\Psi_K}\\
    &\quad{}={} \bra{\Psi_K}
    \pqty{e^{i2\pi J_z/K\hbar}}^k J_x^{(j_n)}
    \pqty{e^{-i2\pi J_z/K\hbar}}^k
    \ket{\Psi_K} \\
    &\quad{}={} \cos(\frac{2\pi k}{K}) \ev{J_x^{(j_n)}} +
    \sin(\frac{2\pi k}{K})
    \ev{J_y^{(j_n)}}.
\end{aligned}
\end{equation}
Taking the sum over $k \in \{0,1,\dots,K-1\}$ and $j \in \mathbf{J}$ on both sides gives $\langle{\sum_{j \in \mathbf{J}} J_x^{(j)}}\rangle = 0$.

Since $e^{-i 2\pi J_z/\hbar K}$ commutes with $J_z^{(j_n)}$, we also have  
\begin{equation}\label{eq:neighbour_sigma_x}
\begin{aligned}
    0 &= \bra{\Psi_K}{J_z^{(j_{n-1})}J_x^{(j_n)}J_z^{(j_{n+1})}}\ket{\Psi_K} \\
    &=
    \bra{\Psi_K}{J_x^{(j_n)}J_z^{(j_{n+1})}}\ket{\Psi_K} \\
    &= \bra{\Psi_K}{J_z^{(j_{n-1})}J_x^{(j_n)}}\ket{\Psi_K}
\end{aligned}    
\end{equation}
for the example states.

Relatedly, with $e^{i2\pi J_z/\hbar K} J_{\pm} e^{-i2\pi/\hbar K} = e^{\mp i2\pi/K} J_{\pm}$ for $J_{\pm}\coloneqq J_x \pm i J_y$, similar steps imply
\begin{equation}
\bra{\Psi_K}{J_{\pm}^2}\ket{\Psi_K} = 0.
\end{equation}
These properties will be useful in later sections.

\subsubsection{Variances of total angular momentum of example states}
The variance $(\Delta J_s)^2 \coloneqq \langle J_s^2 \rangle - \langle J_s \rangle^2$ for $s \in \{x,y,z\}$ are all related to the quantity $\langle J_z^2 \rangle$ for these example states. Starting with $(\Delta J_s)^2 = \langle J_s^2 \rangle$ from $\langle J_s \rangle = 0$, and using $J_+J_- + J_-J_+ = 2\lvert\vec{J}\rvert^2 - 2 J_z^2$, we have
\begin{equation}
\begin{aligned}
    J_x^2 &= \frac{1}{4}\pqty{J_+ + J_-}^2 = \frac{1}{4}\pqty{J_+^2 + J_-^2 + 2\lvert\vec{J}\rvert^2 - 2 J_z^2}, \\
    J_y^2 &= -\frac{1}{4}\pqty{J_+ - J_-}^2 = \frac{1}{4}\pqty{- J_+^2 - J_-^2 + 2\lvert\vec{J}\rvert^2 - 2 J_z^2}.
\end{aligned}
\end{equation}
Since $\langle{J_{\pm}^2}\rangle = 0$, we end up with
\begin{equation}\label{eq:example-states-squared}
    \ev{J_x^2} = \ev{J_y^2} = \frac{1}{2}\ev{\lvert\vec{J}\rvert^2} 
    - \frac{1}{2}\ev{J_z^2}
\end{equation}
for these states.

\subsection{Nondetection by Other Criteria}
Given the above properties, we can now systematically apply each criteria to the example states.

\subsubsection{Spin squeezing inequalities}
Spin squeezing inequalities are GME witnesses that can be expressed in terms of the sum or product of variances of spin operators \cite{teh_reid_2019}. For a spin ensemble $\{j_n\}_{n=1}^N$, these inequalities take the general form
\begin{align}\label{eq: teh&reid ineq}
    (\Delta u)^2 + (\Delta v)^2 &\geq \hbar \min S_B, \\\label{eq:ss-2}
    \Delta u\Delta v &\geq \frac{\hbar}{2}\min S_B,
\end{align}
where $u \coloneqq \sum_{n=1}^N h_{n} J_x^{(j_n)}$, $v \coloneqq \sum_{n=1}^N g_{n} J_y^{(j_n)}$, and $h_n$ and $g_n$ are real numbers. Meanwhile, the set $S_B$ is defined as
\begin{align}
    S_B &\coloneqq \Bigg\{ \left| \sum_{j_\alpha \in \mathbf{J}} h_{\alpha} g_{\alpha} \ev{J_z^{(j_\alpha)}} \right| + \left| \sum_{j_\beta \in \mathbf{J}^\complement} h_{\beta} g_{\beta} \ev{J_z^{(j_\beta)}} \right| \nonumber \\
    &\quad\qquad : \text{$\mathbf{J}$-$\mathbf{J}^\complement$ is a bipartition of $\{j_n\}_{n=1}^N$}
    \Bigg\}.
\end{align}
For the tripartite case, the authors also derived a spin version of an inequality first introduced for continuous variable systems \cite{loock_furusawa}:
\begin{equation}\label{eq:ss-3}
    \min\Bqty{
        \sum_{n=1}^3 B_n, \: \sum_{n=1}^3 S_n
    }
    \geq \hbar \abs{
        \sum_{n=1}^3 \ev{J_z^{(j_n)}}
    }, 
\end{equation}
where
\begin{align}
    B_n &\coloneqq \bqty{
        \Delta\left(
                J_y^{(j_{n'})} + J_y^{(j_{n''})} + g_n J_y^{(j_n)}
            \right)
    }^2  \\
    &\qquad\quad{}+{} \bqty{
        \Delta\left(J_x^{(j_{n'})} - J_x^{(j_{n''})}\right)
    }^2,\nonumber\\
    S_n &\coloneqq 2\Delta\left(J_y^{(j_{n'})} + J_y^{(j_{n''})} + g_n J_y^{(j_n)}
    \right) \\
    &\qquad\quad{}\times{}\Delta\left(
        J_x^{(j_{n'})} - J_x^{(j_{n''})}\right)\nonumber,
\end{align}
and $(n,n',n'')$ is a permutation of $(1, 2, 3)$. Violating any of Eqs.~\eqref{eq: teh&reid ineq}, \eqref{eq:ss-2}, or \eqref{eq:ss-3} witnesses GME.

From Appendix~\ref{apd:example-states-zero-J}, we know that $\langle{J_z^{(j_n)}}\rangle = 0$ for all states in Eq.~\eqref{eq: detected states}, so all inequalities mentioned above become trivial, and no violation occurs. Hence, these inequalities cannot certify the GME of our example states.

Equation~\eqref{eq: teh&reid ineq} was later improved by replacing $S_B$ on the right-hand side of the inequality with \cite{li_gme_2021}
\begin{align}
    \tilde{S}_B \coloneqq &\left\{ \left|\sum_{j_\alpha \in \mathbf{J}} h_{\alpha} g_{\alpha} \ev{J_z^{(j_\alpha)}} \right| + \left| \sum_{j_\beta \in \mathbf{J}^\complement} h_{\beta} g_{\beta} \ev{J_z^{(j_\beta)}} \right| \right. \nonumber\\
    &\qquad {}+{} W^2_{\mathbf{J},\mathbf{J}^\complement} \left. : \text{$\mathbf{J}$-$\mathbf{J}^\complement$ is a bipartition of $\{j_n\}_{n=1}^N$}
    \right\},
\end{align}
where
\begin{align}
    W_{\mathbf{J},\mathbf{J}^\complement} &\coloneqq \sqrt{\Delta^2 u_\mathbf{J} + \Delta^2 v_\mathbf{J} - \left|\sum_{j_\alpha \in \mathbf{J}}h_{\alpha}g_{\alpha}\ev{J_z^{(j_\alpha)}}\right|} \nonumber \\
    &\qquad{}-{} \sqrt{\Delta^2 u_{\mathbf{J}^\complement} + \Delta^2 v_{\mathbf{J}^\complement} - \left|\sum_{j_\beta \in \mathbf{J}^\complement}h_{\beta}g_{\beta}\ev{J_z^{(j_\beta)}}\right|}
\end{align}
for $u_{\mathbf{J}} \coloneqq \sum_{j_{\alpha} \in \mathbf{J}}h_{\alpha}J_x^{(j_{\alpha})}$ and  $v_{\mathbf{J}} \coloneqq \sum_{j_{\alpha} \in \mathbf{J}}g_{\alpha}J_y^{(j_{\alpha})}$,
with $u_{\mathbf{J}^\complement}$ and $v_{\mathbf{J}^\complement}$ similarly defined.
Again with $\langle{J_z^{(j_n)}}\rangle = 0$, the inequality becomes
\begin{align}
    &\ev{(u_\mathbf{J} + u_{\mathbf{J}^\complement})^2} + \ev{(v_\mathbf{J} + v_{\mathbf{J}^\complement})^2} \nonumber\\ 
    &\qquad{}\geq{}\left(\sqrt{\ev{u_\mathbf{J}^2} + \ev{v_\mathbf{J}^2}} - \sqrt{\langle{u_{\mathbf{J}^\complement}^2}\rangle + \langle{v_{\mathbf{J}^\complement}^2}\rangle}\right)^2,
\end{align}
which is also trivially true, so the stronger form of this witness cannot detect the entanglement of the states under consideration.

\subsubsection{Spin-half chain criterion}
Another GME witness proposed for an $N$-partite spin-half chain involves the violation of the inequalities \cite{toth_singlets_2004}
\begin{align}
    \ev{\sum_{n=1}^N \tilde{\sigma}_x^{(j_n)}} &\leq \frac{N}{2}, &
    \sum_{n=1}^N \ev{\tilde{\sigma}_x^{(j_n)}}^2 &\leq \frac{N}{2}
\end{align}
for $\tilde{\sigma}_x^{(j_n)} = \sigma_z^{(n-1)}\sigma_x^{(j_n)}\sigma_z^{(j_{n+1})}$ and $\sigma_z^{(j_0)} = \sigma_z^{j_{N+1}} = \mathbbm{1}$, where $\sigma_s^{(j_n)} \coloneqq 2J_s^{(j_n)}/\hbar$.

From Eq.~\eqref{eq:neighbour_sigma_x}, $\langle{\tilde{\sigma}_x^{(j_n)}}\rangle = 0$ for all states under consideration, so the above inequalities are trivially satisfied. Therefore, they cannot be detected by these criteria.
\begin{widetext}
\subsubsection{Generalized spin squeezing inequalities}
Generalized spin squeezing inequalities were first introduced for spin-half ensembles \cite{toth_ssi_2007, toth_ssi_2009}, and later extended to larger spins \cite{vitagliano_sse_2011, vitagliano_sse_2014}. For an ensemble containing $N$ spin-$j$ particles, they take the form
\begin{align}\label{eq: vitagliano ineq}
    D_1 &= (\Delta J_x)^2 + (\Delta J_y)^2 + (\Delta J_z)^2 - N \hbar^2 j,\\
    D_2^{(qrs)} &= (N - 1)(\tilde{\Delta} J_s)^2 - \left(\ev{\tilde{J}_q^2} + \ev{\tilde{J}_r^2}\right)+ N(N-1)\hbar^2 j^2,\\\label{eq: vitagliano ineq last}
    D_3^{(qrs)} &= (N-1)\bqty{
        (\tilde{\Delta} J_q)^2 +
        (\tilde{\Delta} J_r)^2
    } - \ev{\tilde{J}_s^2} + N(N-1)\hbar^2 j^2,
\end{align}
where $(q,r,s)$ is a permutation of $(x,y,z)$, and
\begin{equation}\label{eq:define-tilde}
    \ev{\tilde{J}_s^2} \coloneqq \ev{J_s^2} - \sum_{n=1}^N \ev{\left(J_s^{(j_n)}\right)^2}
    = 2\sum_{n \neq n'} \ev{J_s^{(j_n)}J_s^{(j_{n'})}},\qquad\qquad 
    \left(\tilde{\Delta}J_s\right)^2 \coloneqq \ev{\tilde{J}_s^2} - \ev{J_s}^2.
\end{equation}
Entanglement is certified by the negativity of any of the following quantities in Eqs.~\eqref{eq: vitagliano ineq}~to~\eqref{eq: vitagliano ineq last}.

Since $\langle{\vec{J}}\rangle=0$ for the example states,
which gives $(\Delta J_s)^2 =  J_s^2$ and $(\tilde{\Delta}J_k)^2 = \langle{(\tilde{J}_k)^2}\rangle$, we have the simplification
\begin{align}
    D_1 &= \ev{J_x^2} + \ev{J_y^2} + \ev{J_z^2} - N \hbar^2 j = \ev{\lvert \vec{J}\rvert^2} - N\hbar^2 j \\
    D_2^{(qrs)} &= (N - 1)\ev{\tilde{J}_s^2} - \left(\ev{\tilde{J}_q^2} + \ev{\tilde{J}_r^2}\right)+ N(N-1)\hbar^2 j^2 \nonumber\\
        &= N\ev{\tilde{J}_s^2} + N(N-1)\hbar^2 j^2 - \sum_{r \in \{x,y,z\}} \ev{\tilde{J}_r^2} \label{eq:D2} \\ &= N\ev{\tilde{J}_s^2} + N(N-1)\hbar^2 j^2 - \ev{\lvert{\vec{J}}\rvert^2} + \underbrace{\sum_{n=1}^N \ev{\lvert{\vec{J}^{(j_n)}}\rvert^2}}_{N\hbar^2 j(j+1)} \nonumber\\
        & = N\ev{\tilde{J}_s^2} + N\hbar^2 j(N j+1) - \ev{\lvert{\vec{J}}\rvert^2}, \nonumber \\
    D_3^{(qrs)} &= (N - 1)\pqty{
        \ev{\tilde{J}_q^2} +
        \ev{\tilde{J}_r^2}
    } - \ev{\tilde{J}_s^2} + N(N-1)\hbar^2 j^2 \label{eq:D3} \\
    &= N\pqty{\ev{\tilde{J}_q^2} + \ev{\tilde{J}_r^2}} + N\hbar^2 j(N j+1) - \ev{\lvert{\vec{J}}\rvert^2}, \nonumber
\end{align}
which depends only on $\langle{\tilde{J}_s^2}\rangle$ and $\lvert \vec{J}\rvert^2$. Using $\lvert \vec{J}\rvert^2 \leq N\hbar^2 j(N j+1)$, we also have convenient lower bounds
\begin{equation}\label{eq:D23-bounds}
    D_2^{(qrs)} \geq N\ev{\tilde{J}_s^2}, \qquad D_3^{(qrs)} \geq N\pqty{\ev{\tilde{J}_q^2} + \ev{\tilde{J}_r^2}}.
\end{equation}
Hence, the nonnegativity of $\langle{\tilde{J}_s^2}\rangle$ for all $s \in \{x,y,z\}$ is sufficient to imply the nonnegativity of $D_2^{(qrs)}$ and $D_3^{(qrs)}$.

We can now compute $D_2^{(qrs)}$ and $D_3^{(qrs)} $ for each of our example states. For the GHZ-like state defined in Eq.~\eqref{eq: detected states}, which for the $N$-partite spin-$j$ chain is written as $
\ket{\text{GHZ}_K} \propto \ket{j,j}^{\otimes N} + (-1)^{\frac{K-1}{2}}\ket{j,-j}^{\otimes N}$, we have
\begin{equation}
\begin{aligned}
    \bra{\text{GHZ}_K} \lvert{\vec{J}}\rvert^2 \ket{\text{GHZ}_K} &= N \hbar^2 j(Nj+1) \\
    \bra{\text{GHZ}_K} {\tilde{J}_s^2}\ket{\text{GHZ}_K} 
    &= \sum_{\mu \in\{+j, -j\}}\left(
        \sum_{m \neq n}
        \bra{j,\mu}J_s^{(j_m)}\ket{j,\mu}
        \bra{j,\mu}J_s^{(j_n)}\ket{j,\mu}
    \right) \\
    &= \begin{cases}
        0, & \text{if $s \in \{x, y\}$}\\
        N(N-1)\hbar^2 j^2 & \text{otherwise.}
    \end{cases}
\end{aligned}
\end{equation}
Since $\bra{\text{GHZ}_K} \lvert{\vec{J}}\rvert^2 \ket{\text{GHZ}_K}-N\hbar^2j = N^2\hbar^2j^2 \geq 0$ and $\bra{\text{GHZ}_K} {\tilde{J}_s^2}\ket{\text{GHZ}_K} \geq 0$ for all $s$, none of the quantities in Eqs.~\eqref{eq: vitagliano ineq}~to~\eqref{eq: vitagliano ineq last} are negative.

We turn to the example states $\ket{\Phi_5}$ and $\ket{\Phi_3}$, which are states of an ensemble of seven and nine spin-half particles, respectively. Since $(J_s^{(j_n=1/2)})^2 = \hbar^2 \mathbbm{1}/4$ for all $s$, we have $
\sum_{n=1}^N \langle({J_s^{(j_n = 1/2)}})^2\rangle = N \hbar^2/4$. Meanwhile, by direct computation, we also have
\begin{equation}
    \bra{\Phi_5}\lvert \vec{J} \rvert^2\ket{\Phi_5} = \frac{35}{4}\hbar^2,\quad
    \bra{\Phi_5}J_z^2\ket{\Phi_5} = \frac{25}{4}\hbar^2,\quad
    \bra{\Phi_3}\lvert \vec{J} \rvert^2\ket{\Phi_3} = \frac{99}{4}\hbar^2,\quad
    \bra{\Phi_3}J_z^2\ket{\Phi_3} = \frac{63}{4}\hbar^2.
\end{equation}
From this, we already have $\bra{\Phi_5}\lvert \vec{J} \rvert^2\ket{\Phi_5}-7\hbar^2/2 \geq 0 $ and $\bra{\Phi_3}\lvert \vec{J} \rvert^2\ket{\Phi_3} - 9\hbar^2/2 \geq 0 $, hence $D_1 \geq 0$ for these two states. From Eq.~\eqref{eq:example-states-squared}, we can further work out
\begin{equation}
    \bra{\Phi_5}J_x^2\ket{\Phi_5}= \bra{\Phi_5}J_y^2\ket{\Phi_5} = \frac{\hbar^2}{2}\pqty{\frac{35}{4} - \frac{25}{4}} = \frac{5}{4}\hbar^2,\quad
    \bra{\Phi_3}J_x^2\ket{\Phi_3} =
    \bra{\Phi_3}J_y^2\ket{\Phi_3} = \frac{\hbar^2}{2}\pqty{\frac{99}{4}-\frac{63}{4}} = \frac{18}{4}\hbar^2.
\end{equation}
This gives us
\begin{equation}
    \bra{\Phi_5}\tilde{J}_s^2\ket{\Phi_5} = \begin{cases}
        -\frac{\hbar^2}{2} &\text{if $s \in \{x,y\}$} \\
        \frac{9}{2}\hbar^2 & \text{otherwise,}
    \end{cases}
    \qquad
    \bra{\Phi_3}\tilde{J}_s^2\ket{\Phi_3} = \begin{cases}
        \frac{11}{4}\hbar^2 &\text{if $s \in \{x,y\}$} \\
        \frac{27}{2}\hbar^2 & \text{otherwise.}
    \end{cases}
\end{equation}
Since $\bra{\Phi_3}\tilde{J}_s^2\ket{\Phi_3} \geq 0$, this already implies that the entanglement of $\ket{\Phi_3}$ cannot be witnessed by this criteria. For $\ket{\Phi_5}$, as $\bra{\Phi_5}\tilde{J}_z^2\ket{\Phi_5} \geq 0$ and $\bra{\Phi_5}\tilde{J}_z^2\ket{\Phi_5} + \bra{\Phi_5}\tilde{J}_s^2\ket{\Phi_5} \geq 0$ for $s \in \{x,y\}$, it is clear that $D_2^{(zxy)}$,  $D_3^{(zxy)}$, and $D_3^{(yzx)}$ will be nonnegative. With further calculations, the other quantities can be found to be
\begin{equation}
    D_2^{(xyz)} = D_2^{(yzx)} = \frac{7}{4}\hbar^2, \qquad D_3^{(xyz)}  = 0.
\end{equation}
Therefore, none of the example states can be detected by the generalized spin inequalities.

\end{widetext}
\subsubsection{Energy-based witnesses}
For various spin models with a governing Hamiltonian $H$, the minimum energies $E^\text{sep}$ achievable by separable states are known \cite{toth_spin_2005}. With that separable bound, a spin ensemble is detected to be entangled when $\ev{H} < E^{\text{sep}}$.

As an example, we consider two specific spin models from the cited paper. In both spin models, $N$ spins are arranged on the vertices of a $d$-dimensional cubic lattice with periodic boundary conditions. Every spin pair connected by an edge is taken to be interacting.

\emph{Heisenberg lattice}. For an anti-ferromagnetic Heisenberg Hamiltonian
\begin{equation}\label{eq:Hamiltonian-H}
    H_\text{H}\coloneqq \sum_{s\in\{x,y,z\}} \sum_{\langle{j_n, j_l}\rangle}\sigma_s^{(j_n)} \sigma_s^{(j_l)}
    +  B \sum_{n=1}^N\sigma_z^{(j_n)},
\end{equation}
where $\langle{j_n,j_l}\rangle$ are interacting spin pairs and $\sigma_s^{(j_n)} \coloneqq 2J_s^{(j_n)}/\hbar$, the separable bound for the energy is
\begin{equation}
    E^{\text{sep}}_\text{H} = \begin{cases}-d N\bqty{
        \frac{1}{8}\pqty{\frac{B}{d}}^2 + 1
    } & \text {if $\left|\dfrac{B}{d}\right| \leq 4$}, \\
    -d N\pqty{
        \abs{\frac{B}{d}} - 1
    } & \text {otherwise.} \end{cases}
\end{equation}
Meanwhile, since $\langle\vec{J}\rangle = 0$ implies $\langle{\vec{\sigma}}\rangle = 0$ for all example states, the last term of Eq.~\eqref{eq:Hamiltonian-H} vanishes. Hence, we are only left with the $\sum_{\langle{j_n,j_l}\rangle}\langle{\sigma_s^{(j_n)} \sigma_s^{(j_l)}}\rangle$ terms.

For $\ket{\text{GHZ}_K}$ and $\ket{\Phi_3}$, notice that they are totally symmetric under any permutation of the individual spins. So, the expectation value $\langle{\sigma_s^{(j_n)} \sigma_s^{(j_l)}}\rangle$ is the same for all interacting spin pairs, and since there are $Nd$ of them,
\begin{equation}
    \sum_{\langle j_n, j_l \rangle}\bra{\Psi_K}\sigma_s^{(j_n)} \sigma_s^{(j_l)}\ket{\Psi_K} = Nd\bra{\Psi_K}\sigma_s^{(j_1)} \sigma_s^{(j_2)}\ket{\Psi_K}
\end{equation}
for these two states.

Computing the expectation value $\langle{\sigma_s^{(j_1)} \sigma_s^{(j_2)}}\rangle$ gives us
\begin{align}
\label{eq: EV on symmetric states}
 \sum_{\langle j_n,j_l \rangle}\bra{\text{GHZ}_K}\sigma_s^{(j_n)}\sigma_s^{(j_l)}\ket{\text{GHZ}_K} &= \begin{cases}
        \frac{Nd}{2} & \text{if $s \in \{x,y\}$,}\\
        0 & \text{otherwise,}
    \end{cases} \\
\label{eq: EV on symmetric states 2}
     \sum_{\langle j_n,j_l \rangle}\bra{\Phi_3}\sigma_s^{(j_n)}\sigma_s^{(j_l)}\ket{\Phi_3} &= \begin{cases}
        \frac{Nd}{8} & \text{if $s \in \{x,y\}$,}\\
        \frac{3Nd}{4} & \text{otherwise.}
    \end{cases}
\end{align}
For the state $\ket{\Phi_5}$, since the only possible cubic lattice with seven spins is a spin chain, we only need to consider the $d=1$ case. Direct computation gives
\begin{align}
\label{eq: EV on Phi_5}
    \sum_{\langle j_n,j_l \rangle}\bra{\Phi_5}\sigma_s^{(j_n)}\sigma_s^{(j_l)}\ket{\Phi_5} &= \begin{cases}
        -\frac{1}{3} & \text{if $s \in \{x,y\}$,}\\
        3 & \text{otherwise.}
    \end{cases}
\end{align}
The positivity of the energies for $\ket{\text{GHZ}_K}$ and $\ket{\Phi_3}$ is evident from Eqs.~\eqref{eq: EV on symmetric states}~and~\eqref{eq: EV on symmetric states 2}. Meanwhile, Eq.~\eqref{eq: EV on Phi_5} also gives
\begin{equation}
    \bra{\Phi_5}H_\text{H}\ket{\Phi_5} = \frac{7}{3} > 0.
\end{equation}
Therefore, since $E^\text{sep}_\text{H} < 0 < \bra{\Psi_K}H_\text{H}\ket{\Psi_K}$, this method is unable to detect the entanglement of all example states.

\emph{$XY$ model}. For the $XY$ Hamiltonian
\begin{equation}
    H_\text{XY} \coloneqq
    \sum_{s\in\{x,y\}} I_s \sum_{\langle{n, l}\rangle}\sigma_s^{(j_n)} \sigma_s^{(j_l)} +
    B \sum_{n=1}^N \sigma_z^{(k)},
\end{equation}
where $\langle{j_n,j_l}\rangle$ are interacting spin pairs and $\sigma_s^{(j_n)} \coloneqq 2J_s^{(j_n)}/\hbar$, the separable bound for the energy is
\begin{align}
\label{eq: E_sep XY}
    E_\text{XY}^\text{sep} = \begin{cases}
    -d N I \pqty{1+\frac{B^2}{4}} & \text {if $b \leq 2$,} \\
    -d N I b & \text {otherwise,}
    \end{cases}
\end{align}
with $I = \max\{|I_x|, |I_y|\}$ and $b = {|B|}/{(Id)}$.

Similar to the case of the Heisenberg lattice, since $E_\text{XY}^\text{sep} < 0$, Eqs.~\eqref{eq: EV on symmetric states}~and~\eqref{eq: EV on symmetric states 2} already rule out the possibility of detecting the entanglement of $\ket{\text{GHZ}_K}$ and $\ket{\Phi_3}$ using this method.

For the remaining state, Eq. \eqref{eq: EV on Phi_5} gives
\begin{align}
    \bra{\Phi_5}H_\mathrm{XY}\ket{\Phi_5} = -\frac{1}{3}(I_x + I_y) \geq -\frac{2I}{3}.
\end{align}
As previously mentioned, the only cubic lattice configuration for seven spins is one-dimensional. Consequently, Eq.~\eqref{eq: E_sep XY} implies that $E_\text{XY}^\text{sep} \leq -7I < \bra{\Phi_5}H_\mathrm{XY}\ket{\Phi_5}$. Therefore, this method also fails to detect the entanglement of $\ket{\Phi_5}$.

\end{document}